\documentclass[aps,pre,twocolumn,notitlepage,superscriptaddress]{revtex4-1}

\usepackage{bm}
\usepackage{graphicx}
\usepackage{latexsym}
\usepackage{revsymb}
\usepackage{amsmath,amssymb}
\usepackage{mathtools}
\usepackage[colorlinks=true,citecolor=blue,linkcolor=blue]{hyperref}
\usepackage{bbding}
\usepackage[caption=false]{subfig} 
\usepackage{float}
\usepackage{subfloat}
\usepackage{graphicx}
\usepackage[section]{placeins}
\usepackage{csquotes}
\usepackage{relsize}
\usepackage{enumitem}
\usepackage[Symbol]{upgreek}
\DeclareGraphicsExtensions{.pdf,.png,.jpg}

\LetLtxMacro{\OldSqrt}{\sqrt}
\newcommand{\ClosedSqrt}[1][\hphantom{3}]{\def\DHLindex{#1}\mathpalette\DHLhksqrt}
\makeatletter
    \newcommand*\bold@name{bold}
    \def\DHLhksqrt#1#2{%
        \setbox0=\hbox{$#1\OldSqrt{#2\,}$}\dimen0=\ht0\relax%
        \advance\dimen0-0.2\ht0\relax
        \setbox2=\hbox{\vrule height\ht0 depth -\dimen0}%
        {\hbox{$#1\expandafter\OldSqrt\expandafter[\DHLindex]{#2\,}$}
        \lower\ifx\math@version\bold@name0.6pt\else0.4pt\fi\box2}
    }
    \renewcommand*{\sqrt}[2][\ ]{\ClosedSqrt[\leftroot{-2}\uproot{1}#1]{#2}\kern0.1em} 
\makeatother

\newcommand{\Lim}[1]{\raisebox{0.5ex}{\scalebox{0.8}{$\displaystyle \lim_{#1}\;$}}}

\begin{document}

\title{Approaching Petavolts per meter plasmonics using structured semiconductors}

\author{Aakash A. Sahai}
\affiliation{Department of Electrical Engineering, University of Colorado Denver, Denver, CO 80204}
\email[corresponding author: ~]{aakash.sahai@ucdenver.edu}

\author{M. Golkowski}
\affiliation{Department of Electrical Engineering, University of Colorado Denver, Denver, CO 80204}

\author{T. Katsouleas}
\affiliation{Department of Electrical Engineering, University of Connecticut, Storrs, CT}

\author{G. Andonian}
\affiliation{Department of Physics, University of California, Los Angeles, CA}

\author{G. White}
\affiliation{Stanford Linear Accelerator Center, Menlo Park, CA}

\author{C. Joshi}
\affiliation{Department of Electrical Engineering, University of California, Los Angeles, CA}

\author{P. Taborek}
\affiliation{Department of Physics, University of California, Irvine, CA}

\author{V. Harid}
\affiliation{Department of Electrical Engineering, University of Colorado Denver, Denver, CO 80204}

\author{J. Stohr}
\affiliation{Stanford Linear Accelerator Center, Menlo Park, CA}


%

\begin{abstract}
A new class of strongly excited plasmonic modes that open access to unprecedented Petavolts per meter electromagnetic fields promise wide-ranging, transformative impact. These modes are constituted by large amplitude oscillations of the ultradense, delocalized free electron Fermi gas which is inherent in conductive media. Here structured semiconductors with appropriate concentration of n-type dopant are introduced to tune the properties of the Fermi gas for matched excitation of an electrostatic, surface ``crunch-in'' plasmon using readily available electron beams of ten micron overall dimensions and hundreds of picoCoulomb charge launched inside a tube. Strong excitation made possible by matching results in relativistic oscillations of the Fermi electron gas and uncovers unique phenomena. Relativistically induced ballistic electron transport comes about due to relativistic multifold increase in the mean free path. Acquired ballistic transport also leads to unconventional heat deposition beyond the Ohm's law. This explains the absence of observed damage or solid-plasma formation in experiments on interaction of conductive samples with electron bunches shorter than $\rm 10^{-13} seconds$. Furthermore, relativistic momentum leads to copious tunneling of electron gas allowing it to traverse the surface and crunch inside the tube. Relativistic effects along with large, localized variation of Fermi gas density underlying these modes necessitate the kinetic approach coupled with particle-in-cell simulations. Experimental verification of acceleration and focusing of electron beams modeled here using tens of Gigavolts per meter fields excited in semiconductors with $\rm 10^{18}cm^{-3}$ free electron density will pave the way for Petavolts per meter plasmonics.
\end{abstract}

\maketitle

\section{Introduction}

Plasmons are quasi-particles that are constituted by modes of collective oscillations of the free electron Fermi gas which is a quantum-mechanical entity inherent in conductive media \cite{Plasmonics-electron-gas, surface-plasmon}. Numerous quantum effects work in unison along with appropriate constituent atoms and ionic lattice structure to bring about the highest, at equilibrium terrestrial density of free electrons in the Fermi gas. This electron gas is delocalized and freely moves about the entire lattice when externally excited. For instance, metals have an equilibrium free electron density as high as $n_0 \rm \simeq 10^{24}cm^{-3}$.

Nano-electromagnetic phenomena that allows nanoscale control over electromagnetic (EM) energy is made possible by this ultra-high electron density which constrains the size or wavelength of plasmons as, 
\begin{align}\label{eq:plasmonic-wavelength}
\begin{split}
	{\rm \lambda_{plasmon}} = 33 ~ ( n_0[10^{24}{\rm cm^{-3}}] )^{-1/2} ~ {\rm nm}.
\end{split}
\end{align}
The ability to nanometrically confine EM energy has resulted in a wide-range of breakthroughs using perturbatively excited optical plasmonics \cite{nanophotonics}. Perturbative plasmons excited by electron microscopy beams have also enhanced imaging at the Angstrom level \cite{ebeam-plasmonics}.

Our work \cite{plasmonics-ieee-2021, spie-2021, pct-2021} has introduced a new physical model of strongly excited plasmonic modes. Strongly excited plasmonics is characterized by spatial amplitude of collective electron oscillations, $\delta$ approaching the plasmonic wavelength, $\rm\delta\simeq \lambda_{plasmon}$. These non-perturbative or large-amplitude plasmons \cite{plasmonics-ieee-2021, spie-2021} are electrostatic due to large-scale electron density displacement $\Delta n_e\simeq n_0$, in contrast with conventional plasmons. Ultimate strength of nonlinear oscillations of free electron Fermi gas is limited by breakdown in coherence. This coherence limit of plasmonic EM fields is put forth in our model to be at least of the order of ``wavebreaking'' limit \cite{wavebreaking-limit} in plasmas,
\begin{align}\label{eq:plasmonic-field}
\begin{split}
	{\rm E_{p}} = \frac{m_ec^2}{e}\frac{2\pi}{\lambda_{\rm plasmon}} \simeq 0.1\sqrt{ n_0[10^{24}{\rm cm^{-3}}] } {\rm PVm^{-1}}.
\end{split}
\end{align}
Electrostatic plasmons sustained by large-scale charge-separation during oscillations of the ultradense, $\rm 10^{24}cm^{-3}$ electron gas are thus capable of unprecedented Petavolts per meter fields. These fields are accessible using techniques such as plasmonic nanofocusing of charged particle beams \cite{plasmonic-nanofocusing} propagating inside tapered tubes to ultrasolid densities.

\begin{figure*}[!htb]
\centering
   \includegraphics[width=0.65\textwidth]{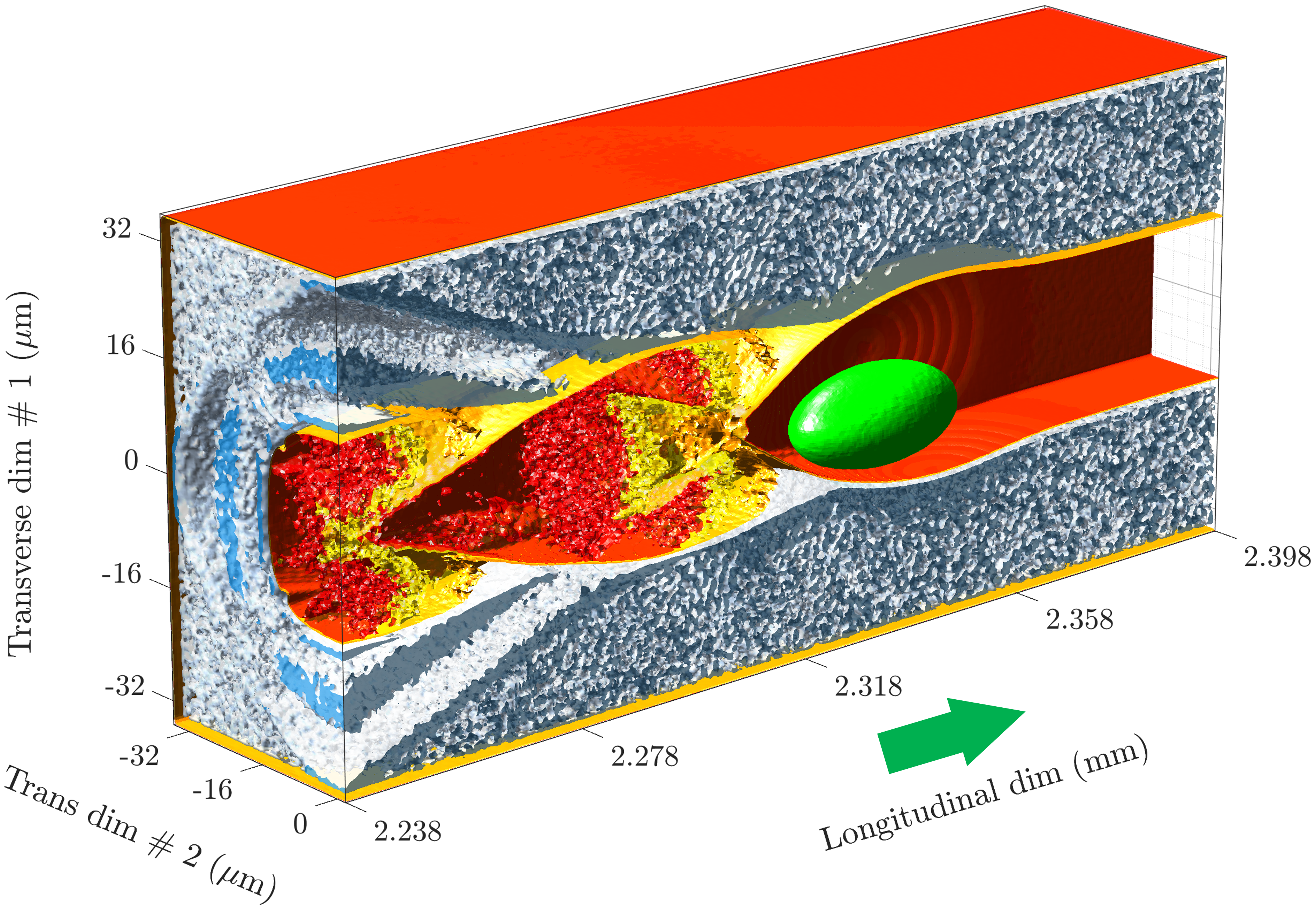}
   \caption{Representation of Fermi electron gas density profile of semiconductor-based surface ``crunch-in'' plasmonic mode excited by a relativistic electron beam (in green) with initial dimensions $\sigma_{z,r}\sim\mathrm{10\mu m}$ propagating (to the right) inside a n-type doped Silicon tube of $25 \times 25\mu m^2$ rectangular cross-section and $\rm 10^{18}cm^{-3}$ conduction band electron density. This is a snapshot  at $\mathrm{2.4 mm}$ ($\sim$ 8ps) of sustained interaction modeled using a three-dimensional (3D) particle-in-cell code.}
\label{fig:3D-semi-crunchin-mode-beam-tube}
\end{figure*}

Access to unprecedented EM fields that are many orders of magnitude higher than other known techniques makes it possible to directly access frontier physics such as ``opening the vacuum'' and ``probing the quantum gravity fabric of spacetime'' discussed in sec.\ref{PV/m-plasmonics-motivation}. In addition, besides various technological applications \cite{pct-2021}, PV/m fields and ultrasolid beams open transformative possibilities for future particle colliders. 

The original work \cite{plasmonics-ieee-2021, spie-2021} introduced relativistic and large-amplitude or nonlinear plasmonic modes that are collisionlessly excited by the collective fields of charged particle beams propagating inside the hollow region of a tube. Large-amplitude plasmons are most effectively excited by ultrashort particle beams. Direct excitation of such plasmons using optical photons ($\rm \lambda_0\gtrsim300nm$) is challenging because their size in typical metals, per Eq.\ref{eq:plasmonic-wavelength} is significantly smaller. More importantly, incoherent energy unavoidably present in the pre-pulse of high-intensity optical lasers ablates the ionic lattice and irreversibly disrupts the electron gas.

As long as the ionic lattice exists, all atomic electrons of conductive condensed matter systems occupy either the valence (bound) or the conduction (free) energy band, in adherence with Bloch's theorem \cite{Bloch-theorem}. Note that the ions are many orders of magnitude heavier than the electrons and therefore, the ionic lattice is only weakly perturbed over few periods of plasmonic oscillations. In accordance with the principles of quantum mechanics, conduction band population that exists without any external excitation and at equilibrium can be engineered to tune the properties of the electron gas.

Using these tunable properties of conductive materials, large-amplitude, electrostatic surface plasmons were modeled \cite{plasmonics-ieee-2021} to be excited by launching an upcoming hundreds of pico-Coulomb (pC) relativistic electron bunch of sub-micron dimensions \cite{gwhite-design,cdr-facet} inside a tube. The predominantly radial ``pancake'' collective fields of the relativistic beam excite a surface plasmon which makes it possible to overcome disruptive instabilities that dominate collision mediated interactions in bulk media.

However, modeling large amplitude plasmons is quite challenging because material properties such as permittivity that hardly diverge from their initial macroscopic average in the perturbative limit have large-scale, highly localized microscopic variations in the nonlinear limit. A kinetic approach coupled with particle-tracking simulations (discussed in sec.\ref{kinetic-model-need}) is therefore introduced in our work for modeling large amplitude plasmons. Specifically, the new physical principle of relativistically induced ballistic transport \cite{plasmonics-ieee-2021} introduced in our work (elucidated below) makes collisionless particle codes suitable for modeling large amplitude plasmons.

In our particle-tracking based computational model, a tube with nanoengineered wall of free electron Fermi gas density $n_t=\rm 2\times 10^{22}cm^{-3}$ was verified to sustain surface ``crunch-in'' plasmonic mode \cite{crunch-in-prab,crunch-in-2015,crunch-in-2015-conf} (where relativistically excited free electrons traverse the surface) with tens of TV/m fields in accordance with Eq.\ref{eq:plasmonic-field}. 

The crunch-in mode of our model is found to be significantly different from the conventional, purely electromagnetic surface plasmons that are transverse magnetic (TM) \cite{Sarkar-review}. While the TM surface plasmon is sustained by perturbative surface electron oscillations, relativistically excited electrons that sustain the surface crunch-in plasmon oscillate across the tube surface and crunch inwards towards the tube axis. Focusing fields of this non-TM, nonlinear surface plasmon \cite{plasmonics-ieee-2021} also enable guiding of the beam through the hollow core of the tube.

Effective excitation of large-amplitude plasmons requires that the dimensions of the beam ($\rm\sigma_r,\sigma_z$) approach those of the plasmon. In \cite{plasmonics-ieee-2021}, wall density $n_t$ and tube radius $r_t=\rm 100nm$ were chosen for a design surface plasmonic wavelength of $\rm \lambda_{plasmon}\simeq250nm$. The dimensions of the electrostatic surface plasmon were thus matched to the sub-micron nanoCoulomb electron beam of bunch length $\rm\sigma_z=400nm$ \cite{gwhite-design} and waist-size $\rm\sigma_r=250nm$ \cite{cdr-facet}. Large collective fields of such dense beams, $n_b\gtrsim10^{20}\rm cm^{-3}$ impart relativistic momentum ($\gamma_{\rm plasmon}\beta_{\rm plasmon}m_ec$) to the free electron Fermi gas which relativistically elongates the plasmon dimensions $\gamma_{\rm plasmon}\lambda_{\rm plasmon}$ and enhances the match. However, accessibility to this combination of bunch charge and beam dimensions may still be several years away.

In this work, to promptly prototype large amplitude plasmonic modes, structured semiconductors are introduced and modeled to sustain large amplitude, electrostatic surface plasmons using a present day beam described in sec.\ref{facet-run-1-beam}. The free electron Fermi gas density in semiconductors can be widely tuned by appropriately choosing the type and concentration of dopant atom embedded in the ionic lattice. Extrinsic semiconductors up to the limit of degeneracy \cite{degenerate-semi} have a tunable conduction band electron density between $10^{13-21} \rm cm^{-3}$ and a corresponding plasmonic wavelength using Eq.\ref{eq:plasmonic-wavelength},
\begin{align}\label{eq:semiconductor-plasmonic-wavelength}
\begin{split}
	{\rm \lambda_{plasmon}} = 3.3 ~ ( n_0[10^{20}{\rm cm^{-3}}] )^{-1/2} ~ {\rm \mu m}.
\end{split}
\end{align}

Semiconductor plasmons are thus highly tunable and allow spatiotemporal match with readily available particle beams. Therefore, by deploying suitably doped semiconductors it becomes possible to immediately prototype and determine the principles underlying electrostatic plasmons that enable PV/m plasmonics. Specifically, n-type extrinsic semiconductors of appropriately doped free electron concentration which allow control over spatial dimensions of plasmons are demonstrated to match with ten micron electron beam. It is also of interest to note that perturbative semiconductor plasmonics \cite{semiconductor-plasmonics} is an active area of research.

However, the operating bunch length for PV/m plasmonics is ideally limited to less than a few tens of micron to avoid affecting the ionic lattice during or after the interaction. In particular, because experiments \cite{Stohr-FFTB-2009} have observed absence of damage or solid plasma formation in conductive samples upon interaction with bunches shorter than $\rm10^{-13}sec$ (see sec.\ref{enabling-principles}).

\begin{figure}[!htb]
\centering
   \includegraphics[width=0.9\columnwidth]{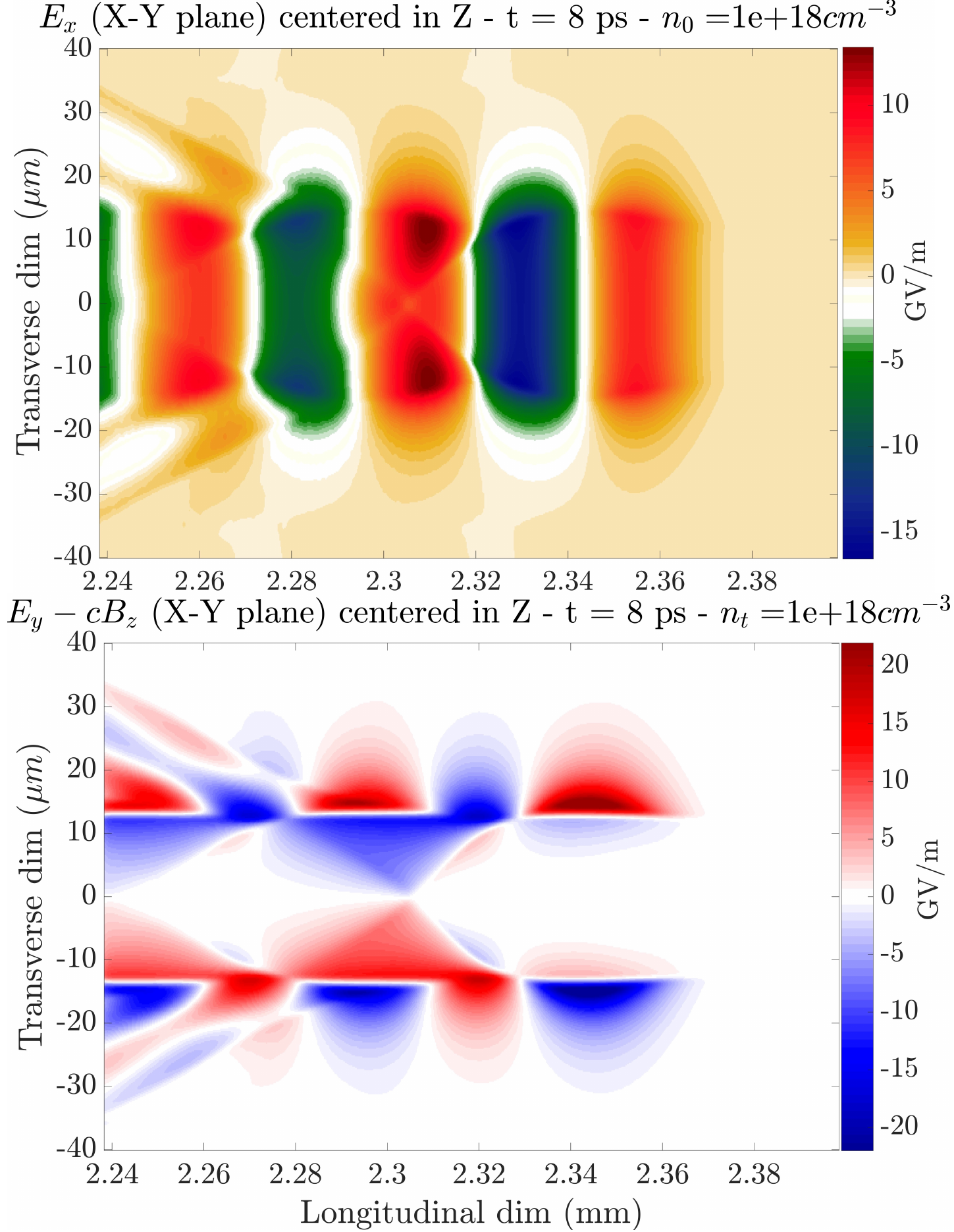}
   \caption{Cross-sections of the accelerating (top panel) and focusing (bottom panel) field profile of the semiconductor-based surface crunch-in plasmonic mode in Fig.\ref{fig:3D-semi-crunchin-mode-beam-tube}}
\label{fig:3D-semi-crunchin-mode-fields}
\end{figure}

Proof of concept of matched excitation of large-amplitude semiconductor surface crunch-in plasmon using readily accessible electron beam is furnished using three-dimensional (3D) kinetic computational model based upon highly parallelized particle-in-cell (PIC) simulation carried out using several thousands of cores of the NSF XSEDE RMACC Summit supercomputer \cite{xsede-rmacc-citation-1,xsede-rmacc-citation-2}. Fig.\ref{fig:3D-semi-crunchin-mode-beam-tube} shows the contours of density profile of the Fermi electron gas to elucidate the mode. The semiconductor tube has a square cross-section of $\rm 25\mu m \times 25\mu m$ and doped to have equilibrium conduction band electron density of $n_t = \rm 10^{18}cm^{-3}$. This tube is excited by launching a currently accessible relativistic ($\rm\gamma_b\sim 10^4$) nC electron beam with dimensions, $\sigma_{z,r}=\mathrm{5\mu m}$. It is to be noted that this is the first model of surface crunch-in plasmon in a rectangular tube. The field profile, longitudinal (top) and focusing (bottom), of the electrostatic, surface crunch-in plasmon excited in the semiconductor tube is in Fig.\ref{fig:3D-semi-crunchin-mode-fields}. Details of the methodology and setup of the simulation in open-source EPOCH code \cite{epoch-pic} are in sec.\ref{kinetic-model-need}.

The tens of GV/m plasmonic fields sustained in $\rm 10^{18}cm^{-3}$ semiconductor tube in Fig.\ref{fig:3D-semi-crunchin-mode-fields} are at the lower end of plasmonic reach. However, even at the lower end experimental investigations stand to unravel the underlying principles of these unexplored modes. Specifically, relativistic momentum acquired by the electron gas under matched excitation leads to the emergence of new phenomena. Plasmonic dynamics is significantly modified besides relativistic dimensions noted above.

Large-amplitude processes of free electron Fermi gas oscillations uncovered in our model \cite{plasmonics-ieee-2021}, namely relativistically induced ballistic electron transport and relativistic ultrafast electron tunneling (see \ref{new-effects-plasmonics}) begin to dominate the plasmonic dynamics. These effects further diverge from conventional plasmonics. In addition to the non-perturbative approach due to nonlinearity where spatiotemporally averaged macroscopic material properties are ineffective, these new phenomena necessitate the kinetic approach which can take into account relativistic and large scale local accumulation of particles which engenders spatially dependent constitutive properties varying over plasmonic timescales.

The large amplitude semiconductor plasmonics effort based on matched excitation with ten micron beam elucidated below is also contrasted against dielectric and metallic tube. This helps immediate experimental verification and delineation of underlying principles and further bolsters the framework of PV/m plasmonics.

In sec.\ref{PV/m-plasmonics-motivation} non-collider possibilities to directly access frontiers of fundamental science using PV/m plasmonic fields are briefly described. Framework of large amplitude plasmonics excited by dense particle beams is elucidated in sec.\ref{framework}, as follows: key enabling advances that make this initiative technologically practicable now in \ref{enabling-principles}, characteristics of free electron Fermi gas in \ref{free-electron}, higher order effects due to the presence of ionic lattice in \ref{ionic-lattice-timescales}, new phenomena due to relativistically oscillating free electron Fermi gas in \ref{new-effects-plasmonics}, disambiguation from dielectrics, optical plasmons and plasma based techniques in \ref{plasmonic-disambiguation}, comparison of electrostatic surface modes against conventional electromagnetic surface mode in \ref{surface-mode-diff}, selected analytical results in \ref{plasmonic-analysis} and dynamic aperture of surface crunch-in modes in \ref{dynamic-aperture}.

In sec.\ref{kinetic-model-need}, kinetic modeling approach is described further. The phase-space of the currently available ten micron, nC relativistic electron beam obtained from beamline simulations is summarized in sec.\ref{facet-run-1-beam}. By porting this phase-space into PIC simulations, effect of large amplitude semiconductor plasmons is estimated with: properties of plasmonic modes excited in different tube dimensions in sec.\ref{tube-radius-scan} and the effect of plasmonic fields on beam transverse phase-space and energy spectra in sec.\ref{plasmonic-effect-beam}. Besides estimation of experimental signatures of fields of semiconductor plasmons on the beam, comparison against dielectric and metal tube provides distinction. Semiconductor plasmons are contrasted against plasmons excited by the same beam in metallic tube used in \cite{plasmonics-ieee-2021} in sec.\ref{semi-vs-metal}. The ongoing sample fabrication efforts are also outlined in sec.\ref{semi-fab}.

Although the proof-of-principle described here utilizes semiconductor plasmonics to match with readily available tens of micron beams, access to ultimate limits of PV/m plasmonics aligns with the ongoing efforts on Mega-Ampere peak current beams discussed in sec.\ref{MA-beam}. This includes compression of nC particle bunches to sub-micron dimensions, bunch length in \ref{length-comp} and waist-size in \ref{waist-comp}, using beamline components. 
\vspace{-5.0mm}

\section{Frontier Fundamental science \\ using PV/m plasmonic fields}
\label{PV/m-plasmonics-motivation}

PV/m plasmonic fields open new pathways in fundamental science. Using PV/m fields for particle acceleration opens up ultra-compact particle colliders accessing PeV particle energies using meter-scale machines. Besides drastic reduction in machine size, solid density beams naturally increase the rate of direct collision between particles and open a new paradigm for luminosity which defines event rates in colliders. Solid-based accelerators provide a more direct path to multi-stage machines and being hollow are suitable for positron acceleration and collisional emittance (phase-space volume of the beam) preservation.

On the other hand, in the near-term PV/m fields open the possibility of non-collider examinations of frontier fundamental science. Below we summarize two such selected searches at the limits of current explorations that exemplify the possibilities.
\vspace{-5.0mm}

\subsection{Opening the vacuum with plasmonic fields: \\ nonlinear quantum electro-dynamics (QED)}
In the non-perturbative limit of QED theory, the rate of spontaneous production of positron-electron pairs directly off the vacuum increases under extreme electromagnetic fields that approach the Schwinger field limit of $\rm E_s = 1.3 \times 10^{18} V/m$ \cite{schwinger-limit}. This rate increases exponentially with the electric field $\rm E_{lab}$ as $\rm exp(-\frac{\pi m_e^2}{eE_{lab}})$ as is typical of tunneling effects. 

Plasmonic fields allow nearing the Exavolts per meter Schwinger field limit and help direct observation of ``opening the vacuum''. Access to extreme fields is based upon the use of plasmonic fields to nano-focus particles beams \cite{plasmonic-nanofocusing}. Plasmonic nanofocusing increases the electric field of particle beams similar to the well known effect in optical plasmonics where tubes with longitudinally tapered radius are used to nano-focus the EM energy coupled into optical plasmons \cite{Stockman-PRL-2004}. 

While there are other proposed schemes to experimentally examine high-field breakdown of vacuum, such as colliding two high-intensity laser pulses or the collision of an ultra-relativistic particle beam with an intense laser pulse, PV/m plasmonics offers certain distinct capabilities. Electromagnetic fields sustained by plasmons have a near-zero group velocity and are localized in space. Therefore, plasmonic fields:
\begin{itemize}[topsep=0pt, itemsep=0.1ex, partopsep=0.3ex, parsep=0.3ex, leftmargin=*]
 \item do not travel with the particle or laser beams, or 
 \item need not exist only in a Lorentz-boosted frame, or
 \item do not exist only momentarily at a collision point
\end{itemize} 
These unique advantages motivate PV/m plasmonics as an attractive alternative to examine nonlinear QED without a collider. Non-collider beams do not have a stringent requirement on their emittance or energy spread and can therefore be utilized for fundamental science.
\vspace{-5.0mm}

\subsection{Probing the fabric of spacetime: \\ quantum-gravity model with ultraenergetic photons}
Theories that attempt to describe gravity using a quantum mechanical framework as well as theory of everything models are considered to be examinable only using astrophysical observation data. 

One of predictions of the string theoretical model that is considered verifiable using astrophysical radiation sources is the non-trivial permittivity to ultra-energetic photons exhibited by vacuum \cite{quantum-gravity}. With the nature of interaction between short wavelength photons and the vacuum, the most appropriate probes of this model are bursty astrophysical sources that produce short pulses of ultra-energetic photons. Therefore, detection of these effects is sorely dependent on observational data.

PV/m plasmonics opens the possibility of bursty TeV or PeV photon production using the nano-wiggler mechanism \cite{plasmonics-ieee-2021}. A nano-wiggler utilizes the extreme focusing fields of plasmonic modes to wiggle the ultrarelativistic electrons of a non-collider beam (low particle count, relatively large emittance and energy spread) that have a relativistic factor, $\rm\gamma_b\simeq 10^{4-7}$. Nanometric wiggling oscillations wavelengths, $\lambda_{\rm osc}$ produce photons of energy, $\rm 2\gamma_b^2 ~ hc/\lambda_{\rm osc}$ ranging from tens of GeV to TeV. Beam particles under the action of PV/m plasmonic fields can thus produce ultra-energetic photons.

Therefore, access to ultraenergetic photons using the nano-wiggler mechanism opens the possibility of examining the predictions of quantum gravity model.
\vspace{-3.0mm}

\section{PV/m plasmonics framework}
\label{framework}

\subsection{Key enablers of PV/m plasmonics}
\label{enabling-principles}

\begin{figure}[!htb]
\centering
   \includegraphics[width=0.75\columnwidth]{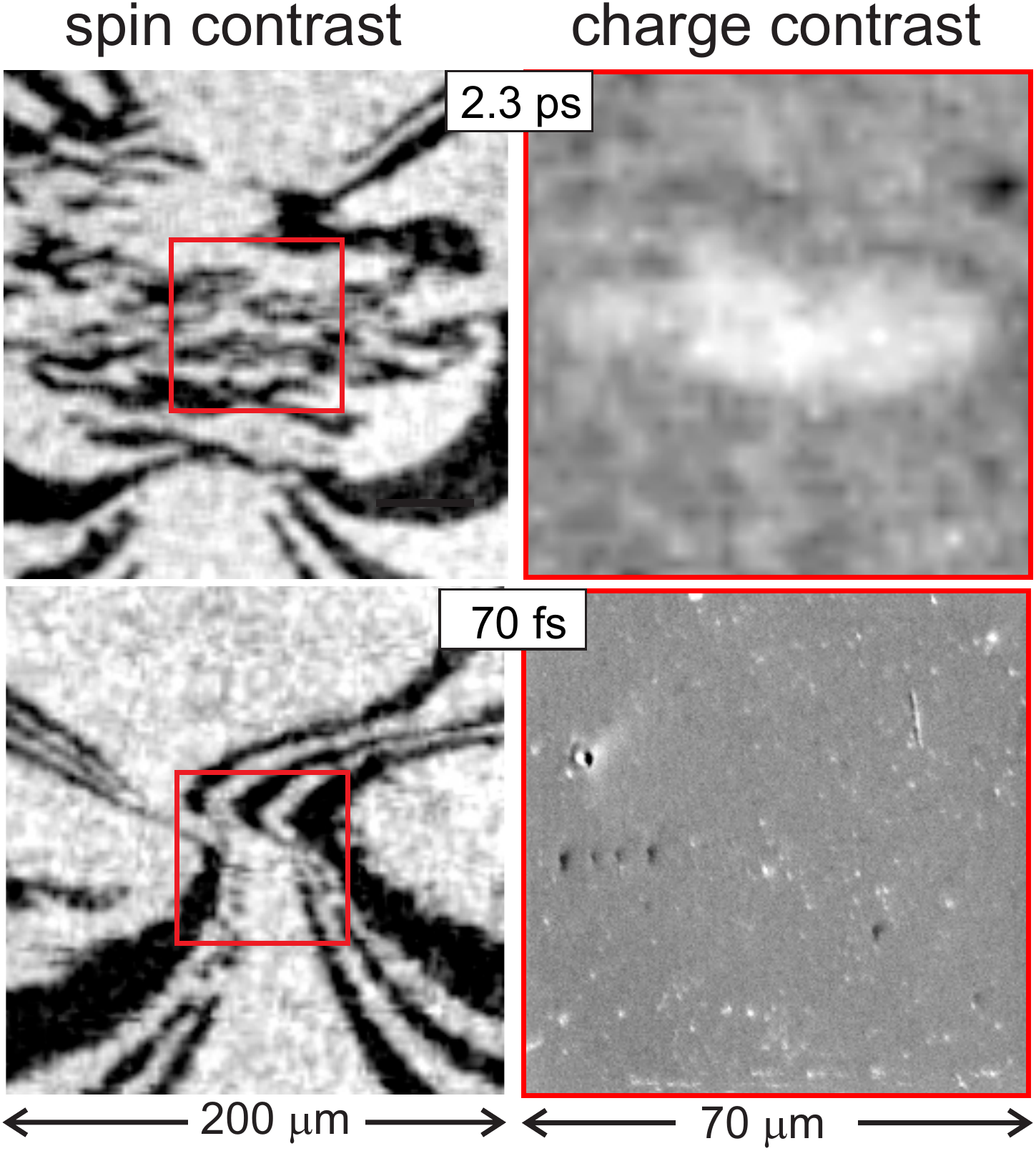}
   \caption{A comparison of the sample images obtained with Scanning Electron Microscope (SEM) with spin polarization (SEMPA), on the left, and conventional SEM of the sample, on the right, zoomed in on the region where the beam hit. It is evident that for electron bunches shorter than $\sigma_{\parallel}\leq \rm 100fs$ there is no topological deformation or solid-plasma formation.}
\label{fig:Stohr-beam-damage}
\end{figure}

Two key experimental advances underpin practical considerations for PV/m plasmonics:
\begin{enumerate}[topsep=2pt, itemsep=0.3ex, partopsep=0.0ex, parsep=0.3ex, wide, labelwidth=!, labelindent=0pt]
\item {\bf No experimentally observed damage of conductive samples for shorter than $20\mu\rm m$ long electron bunch:} Firstly, ultrafast magnetic switching experiments \cite{Stohr-FFTB-2009} observed no damage or plasma formation on a many millimeters-thick conductive magnetized Iron-Cobalt alloy sample using transmission electron microscopy (TEM) after its interaction with a few nC, less than 100fs long, tens of micron waist electron beam. Although no damage was detected using charge-contrast TEM, spin contrast TEM revealed that electron spins had switched over these ultrashort timescales. In contrast, severe damage was observed along with magnetic switching using a 2.3 picosecond long electron beam \cite{Stohr-FFTB-2003}.

Fig.\ref{fig:Stohr-beam-damage} shows a comparison of material damage for the two field pulses of different length and field strength illustrated using spin sensitive transmission electron microscopy in Fig.\ref{fig:Stohr-beam-damage}(a) and conventional transmission electron microscopy in Fig.\ref{fig:Stohr-beam-damage}(b).

The relativistic plasmonic model of large-amplitude modes offers an explanation for this fascinating observation that still remains unexplained. Upon compression of the bunch with a nC charge from 2.3ps to 70fs while the waist-size remains the same, the electric field increases by a factor of 33. This orders of magnitude increase in the beam electric field makes the plasmonic oscillations more relativistic making them ballistic and consequently modifies the heat deposition mechanism. Relativistically augmented mean free path of the Fermi gas do not result in collisional deposition of heat and cannot be explained using the Ohm's law.

The onset of ballistic heat transport by relativistic free electron Fermi gas therefore eliminates the formation of solid-plasma during or after the interaction. This ensures that when a dense charged particle beam less than $\leq\rm 10^{-13} sec$ long induces relativistic oscillations, plasmonic effects dominate and there is no sample damage.

\item {\bf Extreme bunch compression}: Secondly, rapid and ongoing technological innovations have now made it possible to compress a charged particle beam containing $\rm 10^{10}$ particles to sub-micron overall dimensions. Recent beamline simulations \cite{gwhite-design} have demonstrated that a nC electron beam can be compressed to 100nm bunch length within the next decade \cite{cdr-facet}. Similarly, magnetic lens based focusing systems are being prototyped to compress the beam waist-size to hundreds of nm. Effort towards MegaAmpere peak current beams with sub-micron waist sizes is summarized in sec.\ref{MA-beam}.
\end{enumerate}
\vspace{-5.0mm}

\subsection{Free electron Fermi gas: a quantum entity}
\label{free-electron}
The free electron Fermi gas is governed by several key quantum mechanical effects. Periodic structure of the ionic lattice as well as the nature of inter-atomic bonding due to the overlap of orbitals of neighboring atoms is critical to the existence of the Fermi electron gas. Potential of the periodic lattice confines orbital electronic wave-functions such that distinct electronic energy bands are formed in accordance with Bloch's theorem \cite{Bloch-theorem}. Energy bands are near-continuum levels of electron energy states that although themselves partitioned by an infinitesimal quanta of energy,  are separated from other bands by a substantial forbidden energy gap. 

In conductive materials, the conduction band contains a dense gas of free electrons that are not tied to any specific atom in the lattice. In particular, electrons that occupy energy states in the conduction band constitute a free electron gas which follows the Fermi-Dirac statistics. The Pauli's exclusion principle \cite{Pauli-spin-statistics} forbids electron-electron interaction making it possible to form an ultra-dense quantum gas of non-interacting Fermions. 

The Fermi electron gas freely moves about the entire lattice. Upon external excitation, it undergoes collective oscillations \cite{Plasmonics-electron-gas}. Over plasmonic timescales the structure of the ionic lattice may get distorted but does not get disrupted. In our work the electron gas is driven to its coherence limit and sustains extreme collective fields which form the basis of PV/m plasmonics.

\vspace{-5.0mm}
\subsection{Ionic lattice and interband transitions}
\label{ionic-lattice-timescales}

The ultrashort particle beams (less than 100 femtoseconds) suitable for PV/m plasmonics is considerably shorter than the timescale of ion motion. So, the ionic lattice as well as the energy bands exist over the timescales of existence of plasmons. Therefore, plasmonic modes dominate the interaction. 

However, higher order effects under the action of intense fields such as interband transition of the valence band electrons to the conduction band, changes in band structure due to lattice distortion or phonon oscillations etc. have to be taken into account.

The interband transition rate is estimated taking into account the frequency content of the EM pulse (coherent and incoherent content) created by the particle bunch. The method of virtual photons or Weizsacker-Williams method \cite{virtual-photon-1, virtual-photon-2} is used to calculate the field of virtual photon bunch with equivalent frequency components. The half-cycle unipolar EM pulse equivalent to the particle bunch does have a direct current component which however plays no role in estimation of interband transitions. The bunch of virtual photons converts to that of real photons upon interaction of the particle bunch with an external particle or field.

The ten micron long ($\leq 10^{-13}$ second) bunches of our work have a coherent virtual photon frequency spectrum which is restricted to few tens of THz with the bunch length being comparable to THz wavelengths. The equivalent photon energy (tens of milli eV) of THz wavelengths are well below the interband transition energy in semiconductors which are typically an eV. 

So, in this work interband transitions and other higher-order effects are safely neglected. However, in the future when bunch lengths approach submicron dimensions as discussed in sec.\ref{MA-beam} and the beam field increases, our modeling will incorporate high-order effects.

\vspace{-5.0mm}
\subsection{New effects in large-amplitude plasmonics}
\label{new-effects-plasmonics}

{\bf Relativistically induced ballistic transport:} In perturbative plasmonics, oscillations are collisionless because oscillation amplitude is much smaller than the wavelength $\delta \ll \lambda_{\rm plasmon}$ and also the mean free path $\delta < \ell_{\rm mfp}$. In \cite{plasmonics-ieee-2021}, increase in the mean free path with free electron momentum, $\ell_{\rm mfp}\propto p_e$ was identified as a key effect underlying relativistic plasmonics.

The mode of free electron transport in conductive media transitions from the usual Ohmic to ballistic \cite{ballistic-transport} as $\ell_{\rm mfp}$ becomes longer than the thickness of the media. Under ballistic condition, the free electron gas has negligible collisions or scattering with impurities and resultantly the Ohm's law does not apply. It is important to note that, in general, classical transport of carriers is considered under continuously applied external electric field (in the form of terminal voltage) which results in net or average transport of the free electron Fermi gas across a conductive media between the electrodes. 

However, electron transport in plasmonic oscillations is excited by pulsed excitation without any direct external injection of electrons from an electrical contact. Therefore, average electron transport in oscillatory motion is spatially limited to the order of plasmonic wavelength of the specific underlying media.

When large amplitude oscillations are excited such that the Fermi gas electrons gain relativistic momentum, these relativistically oscillating electrons acquire induced ballistic character.  As relativistic plasmonic modes have not been previously studied, our work uncovers and finds this effect to be quite critical for large amplitude plasmonic dynamics. Specifically, for electrons of the Fermi gas oscillating with an average Lorentz factor of $\langle\gamma_e\rangle$, the relativistic mean free path increases as $\langle\gamma_e\rangle^k\times\ell_{\rm mfp}$ where, $k$ is a scaling factor which depends upon the material and other related factors.  

This relativistically induced multifold increase is also experimentally observed with externally incident electron beams and is interpreted to be due to significant reduction in interaction length of electron with an ionic center in the lattice. Typically, the interaction cross-section $\sigma\propto\ell_{\rm mfp}^{-1}\propto\langle\gamma_e\rangle^{-k}$, scales as $\sigma\propto\mathcal{E}^{-2}$ with particle energy $\mathcal{E}$. Therefore, $\ell_{\rm mfp}$ would at least scale as $\gamma_e^2$.

The relativistic ballistic transport model is rooted in the experimental observation of no damage as the beam fields increase with bunch compression from 2.3ps to 70fs, discussed in sec.\ref{enabling-principles}.1. With a 33 times increase in the beam fields and $\langle\gamma_e\rangle$ in experiments, $\ell_{\rm mfp}(\gamma_e)$ is expected to increase by a factor of at least $10^3$. 

{\bf Relativistic tunneling of electron gas which traverses the surface potential:} Another key effect identified in \cite{plasmonics-ieee-2021} is the increase in the probability of ultrafast tunneling across the surface potential under relativistic excitation. The conduction band electrons in metal are capable of tunneling through the surface potential \cite{Fowler-Nordheim}. Especially those electrons with velocities that lie in the tail of the Fermi-Dirac distribution function and whose energy approaches the barrier potential. The rate of tunneling, a near-instantaneous process, depends on  particle energy relative to the barrier potential. 

In perturbative plasmonics, energy of nearly the entire distribution of the excited free electron gas is much smaller than the surface potential. So, tunneling is minimal and inconsequential to the dynamics. However, in large amplitude plasmonics as the electrons are driven to relativistic velocities they become capable of tunneling through the surface, in both directions. This is in addition to their ``above the barrier'' free propagation when their oscillation energy exceeds the surface potential. 

Particularly, as the free electrons are primarily driven transverse to the tube axis, even weakly relativistic oscillations freely oscillate across the surface.
\vspace{-5.0mm}

\subsection{Disambiguation from other mechanisms}
\label{plasmonic-disambiguation}

\begin{enumerate}[topsep=4pt, itemsep=0.3ex, partopsep=0.3ex, parsep=0.3ex, label=\alph*., wide, labelwidth=!, labelindent=0pt]

\item {\bf Dielectric or insulating solids}: Dielectrics or insulating materials have near zero electron density in the conduction band. Although there appear to be phenomenological similarities between the plasmonic modes in conducting media and the electromagnetic modes in an insulting dielectric because both use a hollow solid tube, the physics of each is quite distinct.  

Whereas plasmonic modes are supported by collective oscillations of free electron Fermi gas, dielectric modes are supported by polarization currents generated by the distortion of the electron cloud bound to the ions in the lattice. Plasmonic oscillations are out of phase with the drive electric field while dielectric polarization currents are not. In dielectrics, polarization currents excite Cherenkov radiation which supports the acceleration mechanism. In large-amplitude plasmonics, electrostatic plasmonic modes supported by large-scale charge separation between the free electron Fermi gas and ionic lattice sustain ultra-high gradients.

As a result, the dispersion relation of each type of wave is quite different:  
\begin{itemize}[topsep=0pt, itemsep=0.3ex, partopsep=0.3ex, parsep=0.3ex,leftmargin=*]
	\item $\omega = \frac{\omega_{\rm plasmon}}{\sqrt{2}}=\sqrt{\frac{4\pi n_0e^2}{2m_e}}$ for a plasmonic surface wave, 
	\item $\omega = k~c/n_{\rm d}$ for a dielectric wave, where $n_{\rm d}$ is the index of refraction of the dielectric.
\end{itemize}

\begin{figure}[!htb]
\centering
   \includegraphics[width=\columnwidth]{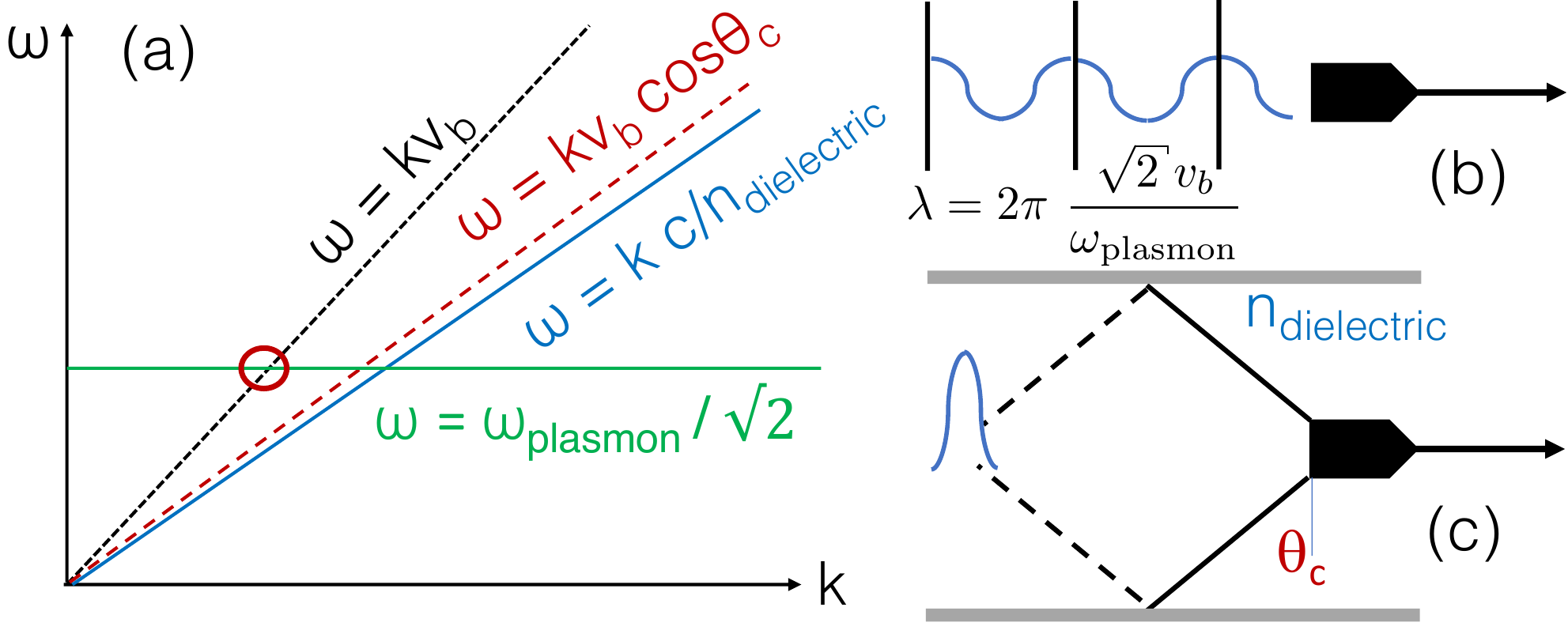}
   \caption{Dispersion relation for surface plasmon and dielectric modes in (a) and their overlap with the drive particle bunch which determines the spatial and temporal profile, (b) of the surface plasmon mode, (c) of the dielectric mode. }
\vspace{-3.0mm}
\label{fig:dispersion-plasmonic-dielectric}
\end{figure}

The resulting mode structure can be predicted from the intersection of the dispersion relations and the driver disturbance at $\omega = k v_b$ (where, $v_b=c\beta_b$ is the beam velocity) shown in Fig.\ref{fig:dispersion-plasmonic-dielectric}a.  From this we see that the surface plasmonic mode has but one wavenumber and frequency and gives rise to planar wave fronts and sinusoidal oscillations as seen in Fig.\ref{fig:dispersion-plasmonic-dielectric}b.  

On the other hand, the dielectric dispersion intersects the beam disturbance at the Cherenkov angle $cos(\theta_c) = 1/n_{\rm d}$ at every $\omega$ and $k$.  This supports the familiar Cherenkov cone type wavefront structure common in the wake of supersonic jets and illustrated in Fig.\ref{fig:dispersion-plasmonic-dielectric}c. This shock-like cone propagates out to the outer conducting wall boundary of the dielectric whereupon it is reflected back toward the axis, giving rise to an accelerating spike on the axis as in Fig.\ref{fig:dispersion-plasmonic-dielectric}c. In order to sustain fields in dielectrics, it is critical to trap the Cherenkov radiation using an outer metallic wall. 

It is clear from this picture that these two modes differ considerably. In the first, the accelerating field is nominally sinusoidal (in the linear regime) and the location of the accelerating peak is controlled by the free electron Fermi gas density in the tube. In the second, the accelerating field is a single spike and its location is determined by the outer radius defined by metallic deposit and the dielectric constant of the tube. 

Experiments have demonstrated that in a dielectric when the bound electrons are freed up by EM fields of intense Cherenkov radiation to transition into the empty conduction band of the dielectric, the electromagnetic wave is rapidly shielded and damped by these free electrons which puts an end to dielectric acceleration \cite{dielectric-damping}.
	
\item {\bf Optical plasmons}: In conventional plasmonics, optical femtosecond laser is used to excite weakly-driven plasmons. Such optical plasmons that have also been proposed as a pathway for particle acceleration \cite{optical-plasmon} are perturbative unlike in our work.
	
\item {\bf Solid plasmas and crystal channeling}: Solid plasmas do not exist at equilibrium and are created by ablation of solids such as by heating the solid with a high-intensity optical (near-infrared) laser. In solid plasmas the individual ions are uncorrelated as there is no ionic lattice. Moreover, solid plasmas do not have an energy band structure. Due to the highly randomized nature of ions in plasmas, the properties of plasma oscillations significantly diverge from that of plasmonic modes especially at higher densities. The individual atoms that have to be ionized to form a plasma result in an electron density that solely depends on the neutral density and the ionization fraction. Also, electron-electron collisions which increase with electron density in a gaseous plasmas are minimized for a quantum mechanical electron gas in adherence with Pauli's exclusion principle. Solid plasmas are utilized for ion acceleration in addition to being proposed for schemes such as channeling acceleration of positively charged particles \cite{solid-plasma}. 

A charged particle injected along the crystallographic axis of a mono-crystalline solid can undergo a series of glancing  collisions or deflections (field-mediated) with periodic row of ions in the lattice. If these interactions with individual ions are capable of trapping the charged particles within the inter-planar space, this phenomena is referred to as crystal channeling \cite{channeling}. Because the basis of energy exchange process is individual interaction of charged particles with the ionic sites, this process is entirely different from plasmonics where energy is exchanged between charged particles and collective fields of Fermi electron gas oscillations.

\end{enumerate}

\vspace{-5.0mm}
\subsection{Purely electromagnetic vs. electrostatic \\ surface modes}
\label{surface-mode-diff}

Surface modes are critical for external excitation of modes in solid media. It is vital to mitigate direct sustained collisions between the particle or photon beam and the ionic lattice in order to prevent complete disintegration of the beam. Direct interaction between particles and the ionic lattice not only result in energy loss of beam particles by random scattering but also disrupts the beam as an entity, through a wide-range of collective instabilities. These instabilities include filamentation, hosing amongst others and can rapidly grow to completely disrupt the beam. 

Large amplitude plasmonics thus utilizes surface plasmons which are excited as the beam propagates through the core region of a tube that is surrounded by conductive walls of desired free electron density.

The conventional surface plasmon (optical or microscopy beam excited) utilizes TM mode \cite{Sarkar-review} which is a purely electromagnetic mode. The TM mode subtends zero focusing forces on particle beam propagating inside the channel \cite{Katsouleas-PRL-1998}. In fact, conventional metallic cavities that form the basis of modern particle particle accelerators and light sources also utilize the TM mode sourced by radiofrequency wall currents. However, the TM mode field structure excited by a beam that is not perfectly aligned to the axis of the cavity produces large deflecting forces. Besides metallic rf cavities, experiments on excitation of the purely electromagnetic TM mode by a charged particle beam in a hollow tube of gaseous plasma have also confirmed these characteristics \cite{Positron-hollow-channel-2016}.
 
On the other hand, the surface crunch-in mode \cite{crunch-in-prab,crunch-in-2015,crunch-in-2015-conf} being high nonlinear violates these conditions imposed by the purely electromagnetic TM mode. In contrast with the TM mode, the  electrostatic surface crunch-in plasmonic mode sustains large focusing forces. These fundamental differences prevent direct comparison between the TM and surface crunch-in modes. 

When the free electron Fermi gas is driven by beam fields larger than tens of MV/m, the free electron gas gains relativistic momentum. With relativistic momentum, the kinetic energy of plasmonic oscillations increases beyond the surface potential and the oscillating electron gas oscillate across the surface \cite{plasmonics-ieee-2021, spie-2021}. In an enclosed geometry, the oscillating electrons that cross over from the surrounding surface into the vacuum experience a restoring force due to mutual Coulomb repulsion.

It is important to note while the electrostatic surface crunch-in mode makes it possible to access coherence limited fields per Eq.\ref{eq:plasmonic-field}, the particle bunches do not directly interact with the background ionic lattice. Due to this property both negatively charged as well as positively charged particles can be accelerated equally well. Apart from the injection of an electron beam, it is also possible to inject ultrashort bunches of positron \cite{Sahai-positron}, positive or negative muons \cite{Sahai-muon}.

\vspace{-5.0mm}
\subsection{Analytical model of surface crunch-in plasmon}
\label{plasmonic-analysis}

The kinetic model of large amplitude plasmons developed from first principles in \cite{plasmonics-ieee-2021} is outlined here. Analytical equation, Eq.\ref{eq:surface-wave-equation} obtained from the first principles models electrons undergoing radial surface crunch-in oscillations. Here, $r_t$ is the tube radius, $n_t$ is the Fermi gas electron density in the tube walls, $Q_b$ is the beam charge, $r_0$ is the initial equilibrium radial position of an electron, $r$ is its instantaneous radial position, $\omega_{p}(n_t)=\sqrt{(4\pi e^2/m_e) ~ n_t}$ (cgs units are used here) is the bulk plasmon frequency, $\gamma_e$ is the relativistic factor acquired by the oscillating Fermi gas electron. 

The Gaussian electron beam that excites the Fermi electron gas has a waist size $\sigma_r$ and bunch length $\sigma_z$, the peak beam density is $n_{b0}=N_b/(2\pi\sqrt{2\pi}\sigma_r^2\sigma_z)$ where $N_b$ is the number of particles in the bunch, relativistic parameters are $\gamma_b, \beta_b$, the instantaneous position in a longitudinal coordinate co-moving with the particle beam is $\xi=c\beta_bt - z$ and $\mathcal{F}$ is the shape function of the continually focused electron beam. 

The kinetic equation of the instantaneous position of the Fermi gas electrons constituting plasmonic oscillation underlying the surface crunch-in mode quoted from \cite{plasmonics-ieee-2021} is,
\begin{align}\label{eq:surface-wave-equation}
\nonumber \frac{\partial^2 r}{\partial \xi^2} & + \frac{1}{2}\frac{\omega_{p}^2(n_t)}{\gamma_e c^2\beta_b^2} ~ \frac{1}{r} ~ \left[ (r^2 - r_t^2) ~ \mathcal{H}(r-r_t) - (r_0^2 - r_t^2)  \right] \\
\nonumber & + \frac{\partial^2 r}{\partial \xi^2} \biggm\lvert_{r=r_t} \mathcal{H}({\rm sgn}[Q_b](r-r_t )) \\
& = - {\rm sgn}[Q_b] ~ \frac{\omega_{p}^2(n_t)}{\gamma_e c^2\beta_b^2} ~ \frac{n_{b0}(\xi)}{n_t} ~ \int_{0}^{r} dr ~ \mathcal{F}(r,z,\xi)
\end{align}

Existence condition of the crunch-in mode under finite wall width $\Delta w$ ensures that all of the tube electrons are not ``blown out'' to escape the lattice ionic potential,
\begin{align}\label{eq:crunch-in-condition}
\begin{split}
n_t\pi\left[ (r_t + \Delta w)^2 - r_t^2 \right] >  n_{b0} ~ \sigma_r^2.
\end{split}
\end{align}

To avoid trajectory crossing at the radial maxima of outward oscillating electrons $r_m$, all the electrons located between $r_t$ and $r_m$ collectively move and bunch together to form a sheath at $r_m$ while maintaining their initial ordering. This maximum radial displacement of outward excursion is obtained from Eq.\ref{eq:surface-wave-equation} for $r_t \gtrsim \sigma_r$ (distinct from flat-top beam of original work \cite{plasmonics-ieee-2021} where, $\sigma_r > r_t$),
\begin{align}\label{eq:crunch-in-max-r}
\begin{split}
r_m - r_t = r_t ~ \left(\quad \left[ 1 + \frac{1}{\pi}\frac{n_{b0}}{n_t} \left(\frac{\sigma_r}{r_t}\right)^2  \right]^{1/2} - 1 \right).
 \end{split}
\end{align}

The net charge of electron rings in the radially inward crunch-in phase of oscillations within an infinitesimal slice thickness, $dz$ at the longitudinal point of maximum compression $\xi=\xi_{r-min}$ using Eq.\ref{eq:crunch-in-max-r} is,
\begin{align}\label{eq:charge-collective-rings-step}
\begin{split}
\delta Q&_{\rm max}(\xi_{r-min})  = - e ~ n_t ~ \pi (r_m^2 - r_t^2) ~ dz.
\end{split}
\end{align}

The radial electric field due to this collective crunch-in of all the free electrons between $r_m$ and $r_t$ to a radius of $r_t/\alpha$ (where $\alpha$ depends upon the mode amplitude) is estimated using Gauss's law in the $r_t \gtrsim \sigma_r$ regime,
\begin{align}\label{eq:crunch-in-charge-radial-field-SI}
\begin{split}
E_{t-r} & = -\frac{2\alpha}{(2\pi)^{3/2}} \frac{eN_b}{r_t\sigma_z} = \frac{- \alpha ~ 11.4 ~ Q_b[{\rm nC}]}{\sigma_z{\rm [10\mu m] r_t[10\mu m]}} {\rm\frac{GV}{m}}.
\end{split}
\end{align}

The crunch-in wavelength is itself proportional to the tube radius $r_t$ (evident from Fig.\ref{fig:facet-ii-run1-tube-radius-scan}). For a large tube $r_t\gg c/\omega_{p}(n_t)$, the crunch-in mode tends to the TM mode ($(r_m-r_t) \ll r_t$) with $\lambda_{\rm crunch-in}\rightarrow \infty$. On the other hand, for bulk solid media where $r_t\rightarrow0$ it is $\lambda_{\rm crunch-in} \simeq 2\pi c/\omega_{p}(n_t)$. Therefore, semi-empirical wavelength of the surface crunch-in plasmon is,
\begin{align}\label{eq:crunch-in-wavelength}
\begin{split}
\lambda&_{\rm crunch-in} = \sqrt{\langle\gamma_e\rangle} \frac{2\pi c}{\omega_{p}(n_t)} ~ \left[ \frac{r_t}{2\pi c/\omega_{p}(n_t)} + 1\right].
\end{split}
\end{align} The phase of relativistically corrected wavelength during which the excited free electrons crunch-in is $\kappa$, typically $\leq0.1$. The average relativistic factor of plasmonic oscillation $\langle\gamma_e\rangle = \sqrt{1 + \left(\frac{\Delta p_r}{m_ec}\right)^2 }$ is estimated using the force of the beam radial fields, $F_{\rm beam}$ that imparts relativistic radial momentum $\frac{\Delta p_r}{m_ec} = \frac{\omega_{p}^2(n_t)}{c^2} ~ \frac{1}{r_t} \left(\frac{n_{b0}}{n_t}\right) \left(\frac{\sigma_r^2\sigma_z}{2\pi}\right)$ to the free electron Fermi gas. Using Panofsky-Wenzel theorem ${\rm\Lim {\Delta\rightarrow0}}E_{t-r}/\Delta \xi = E_{t-z}/(2\pi\Delta r)$ \cite{Panofsky-Wenzel} where $\Delta r=r_m-r_t$ and $\Delta \xi = \kappa ~ \lambda_{\rm crunch-in}$, the maximum longitudinal electric field $E_{t-z}$ is estimated under $r_t \gtrsim \sigma_r$,
\begin{align}\label{eq:crunch-in-charge-acc-field-SI}
\nonumber E&_{t-z} = \frac{-2e ~ \alpha\kappa^{-1}}{(2\pi)^{1/4}} \frac{\sqrt{N_b n_t r_t}}{\sigma_z} ~ \frac{\sqrt{1 + \frac{1}{\pi} \left(\frac{n_{\rm b0}}{n_t}\right)\frac{\sigma_r^2}{r_t^2}} - 1}{1 + r_t/\lambda_p} \\
\nonumber & = -5.2~\frac{\alpha}{\kappa} ~ \frac{\sqrt{ Q_b[{\rm nC}] ~ n_t[10^{18}{\rm cm^{-3}]} ~ r_t \rm [10\mu m]}}{{\sigma_z \rm [10\mu m]}} \\
&\qquad \qquad \times \frac{\sqrt{1 + \frac{1}{\pi}\left(\frac{n_{\rm b0}}{n_t}\right) \frac{\sigma_r^2 {\rm [10\mu m]} }{r_t^2 {\rm [10\mu m]} }} - 1}{1 + r_t/\lambda_p} ~ {\rm \frac{GV}{m}.}
\end{align}
\vspace{-7.5mm}

\subsection{Dynamic aperture interpretation of \\electrostatic surface modes}
\label{dynamic-aperture}
From the Panofsky-Wentzel theorem \cite{Panofsky-Wenzel} applied to purely electromagnetic surface modes such as the TM mode at a given frequency it follows that when the longitudinal field of a charge located inside a structure increases with miniaturization as $1/a^2$, where $a$ is the ``fixed'' transverse dimension of a tube ($r_t$ at equilibrium), the transverse field amplitude increases as $1/a^3$.  These purely electromagnetic transverse fields are unfavorable and lead to head-tail and beam breakup (BBU) instabilities that severely limit the amount of charge that can be accelerated in miniaturized devices. Moreover, transverse misalignment between the beam axis and the axis of symmetry of the structure excites higher-order modes that deflect the beam off-axis.

On the other hand, this scaling can be overcome with ``dynamic'' transverse aperture, $a(\xi)$ as is the case in  electrostatic surface crunch-in mode. The head of an electron beam drives the free electrons Fermi gas outwards which as part of its oscillation trajectory goes across the surface and re-converges on the axis behind the driver.  The aperture behind the beams $a(\xi_{\rm min})$, can be vanishingly small and result in large accelerating fields, while the aperture seen by the head of the beam ($r_t \gtrsim \sigma_r$) and associated with the transverse fields is much larger. The dynamic aperture of electrostatic modes can overcome the scaling limitations that otherwise limit purely EM modes. 
\vspace{-3.0mm}

\section{Large-amplitude semiconductor plasmonics: computational model}
\label{Semi-plasmonics-model}

\subsection{Kinetic modeling of large-amplitude plasmonics \\and simulation setup}
\label{kinetic-model-need}

The relativistic and nonlinear plasmonic modes excited in conductive materials need to solve the full set of Maxwell's equation unlike the free-space (zero charge and current densities, $\rho_e=J_e=0$) assumption behind purely electromagnetic solver. Nonlinear plasmonic modes that are strongly electrostatic do not lend themselves to being modeled using purely electromagnetic codes. Conventional optical plasmons are typically modeled using Finite-Difference-Time-Domain (FDTD) method where the perturbative electron oscillations are simply approximated using constitutive parameters. The FDTD approach cannot be used when the trajectories of collective electron oscillations attain amplitudes which are a significant fraction of  plasmonic wavelength resulting in large-scale charge density displacement. Large density displacement leads to highly localized, rapid spatial variation of material properties such as permittivity which necessitate a non-perturbative approach.

Our work has adopted the kinetic approach along with particle-tracking Particle-In-Cell (PIC) computational modeling of collective oscillations of the free electrons Fermi gas to account for the nonlinear characteristics of strongly electrostatic plasmons. Specifically, the PIC methodology incorporates charge and current densities by tracking representative particles within the FDTD solver to calculate the net electromagnetic fields. In addition, it also calculates the effect of the updated fields on the charge and current densities and accordingly pushes particles under the action of the net EM fields obtained using FDTD completing the loop. This approach is based upon collisionless nature of relativistic oscillations of the Fermi gas stemming from relativistically induced ballistic electron transport explained in sec.\ref{new-effects-plasmonics}. In this methodology, no constitutive parameters are initialized to model materials. Nonetheless, initialization and self-consistent evolution of electron charge and current densities implicitly accounts for effects that the constitutive parameters represent in an averaged sense. 

However, as relativistic oscillations of the free electron Fermi gas have been experimentally observed to go beyond the Ohm's law, the long-term evolution of these oscillations and the effect on ionic lattice requires further development. Specifically, conductivity which is a critical constitutive parameter based upon average electron-ion collisions needs further experimental investigations to be properly understood in the relativistic plasmonic regime.

The 3D PIC simulation in Fig.\ref{fig:3D-semi-crunchin-mode-beam-tube} that models the semiconductor surface crunch-in plasmonic mode is setup over a cartesian spatial grid of $\rm 160\times80\times80\mu m^3$ resolved with 200nm cubic cells. The free electron Fermi gas density of $n_t = \mathrm{10^{18} cm^{-3}}$ in the n-type doped semiconductor tube of square cross-section is modeled using 4 particles per cubic cell with fixed ions. The length of a side of the square cross-section is, $r_t = \mathrm{12.5\mu m}$ and wall thickness $\Delta w \mathrm{= 25\mu m}$. Realistic electron beam profile to enable experimental verification (described in sec.\ref{facet-run-1-beam}) with peak density $n_{b0} = \mathrm{10^{18} cm^{-3}}$; waist-size $\sigma_r \mathrm{\simeq5 \mu m}$ ($r_t=2.5\times\sigma_r$) and bunch length $\sigma_z \simeq \mathrm{10 \mu m}$ is initialized with 2 particles per cell. The box copropagates with the ultrarelativistic beam, $\gamma_b\simeq2\times10^4$. Absorbing boundary conditions are used for both fields and particles.

\begin{figure}[!h]
\centering
   \includegraphics[width=0.95\columnwidth]{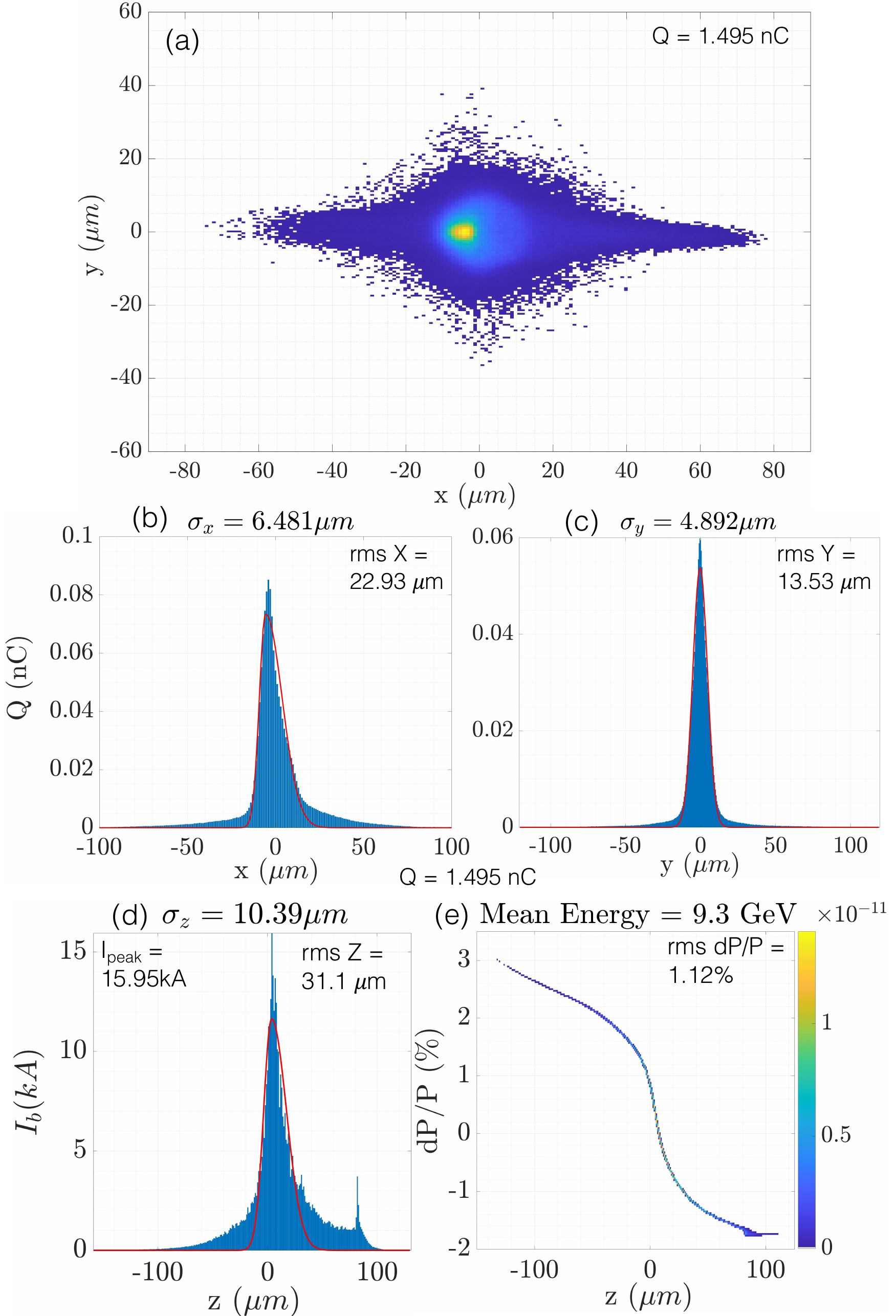}
   \caption{Simulated phase-space slices of early stage FACET-II electron beam: (a) beam transverse profile, (b,c) transverse spatial distribution, (d) longitudinal spatial distribution, (e) longitudinal momentum phase-space. Estimated standard deviation ($\sigma$) and root-mean-square (rms) of the spatial distribution are obtained from a Gaussian fit.}
\vspace{-5.0mm}
\label{fig:facet-ii-run1-phase-spaces}
\end{figure}

The parameters of the crunch-in mode observed from 3D PIC simulation in a tube with square cross-section are in good agreement with our kinetic model (of sec.\ref{plasmonic-analysis}) in cylindrical symmetry. The crunch-in wavelength $\lambda_{\rm crunch-in}$ per Eq.\ref{eq:crunch-in-wavelength}, $\rm\langle\gamma_e\rangle\times 45\mu m$ agrees well with the $\rm 50\mu m$ observed in Fig.\ref{fig:3D-semi-crunchin-mode-fields}. Eq.\ref{eq:crunch-in-charge-radial-field-SI} (Eq.\ref{eq:crunch-in-charge-acc-field-SI}) predicts a peak focusing field (accelerating field) of around $\rm E_{\rm t-r} \simeq 13.7\alpha ~ GV/m$ ($\rm E_{\rm t-z} \simeq 9.3\alpha ~ GV/m$ with $\kappa=0.1$) which agrees well with Fig.\ref{fig:3D-semi-crunchin-mode-fields}, top panel $\rm E_{\rm t-r} \simeq 20 ~ GV/m$ (Fig.\ref{fig:3D-semi-crunchin-mode-fields}, bottom panel $\rm E_{\rm t-z} \simeq 15 ~ GV/m$). 

The 2.5D PIC (2 spatial and 3 velocity dimensions) simulation in sec.\ref{tube-radius-scan},\ref{plasmonic-effect-beam},\ref{semi-vs-metal} utilize cartesian spatial grid with 100nm square cells. The particles of free electron Fermi gas is modeled with 9 macro-particles per cell and the particles of the beam with 4 macro-particles per cell. Beam profile is initialized as per simulated transverse and longitudinal beam phase-spaces in sec.\ref{facet-run-1-beam}.

\vspace{-5.0mm}
\subsection{Phase-space of a readily accessible\\ ten micron, nC beam}
\label{facet-run-1-beam}

Our kinetic simulations utilize parameters of currently available beam at FACET-II facility directly from a beam dynamics simulation. This start-to-end beam dynamics simulation incorporates the FACET-II electron injector, linac and compression systems. The injector comprises a high-gradient copper cathode rf gun, 125 MeV s-band linac and double-bend system to inject into the main s-band linac. The emittance of the 2nC bunch injected from the gun is 5 mm-mrad. This section of beamline is modeled using the General Particle Tracer (GPT) tracking code \cite{GPT-code}, which accounts for 3D space charge as well as rf cavity wakefields in the s-band structures. 

Particles from GPT simulation are passed to the Lucretia code \cite{lucretia-code} which models the rest of the beamline. This includes three additional s-band acceleration sections and three stages of bunch compression at 335 MeV, 4.5 GeV and 10 GeV. Longitudinal and transverse wakefields are modeled in the s-band rf structures. Incoherent (ISR) and coherent (CSR) synchrotron radiation effects \cite{csr-talman} are modeled in the compression chicane bends while longitudinal space charge effect is modeled through the entire accelerator chain. 

From simulations, at a minimum final beta function (distance over which focussed beam diverges to two times the area at its focal waist) of 5 cm (at 10 GeV), the minimum vertical beam waist-size is found to be about $\rm 4 \mu m$, and the horizontal spot size depends on the final bunch length (due to CSR emittance degradation in the final bunch compressor). Fig.\ref{fig:facet-ii-run1-phase-spaces} shows particle distributions for transverse and longitudinal phase space slices of early stage FACET-II electron beam.

\begin{figure*}[!htb]
\centering
   \includegraphics[width=\textwidth]{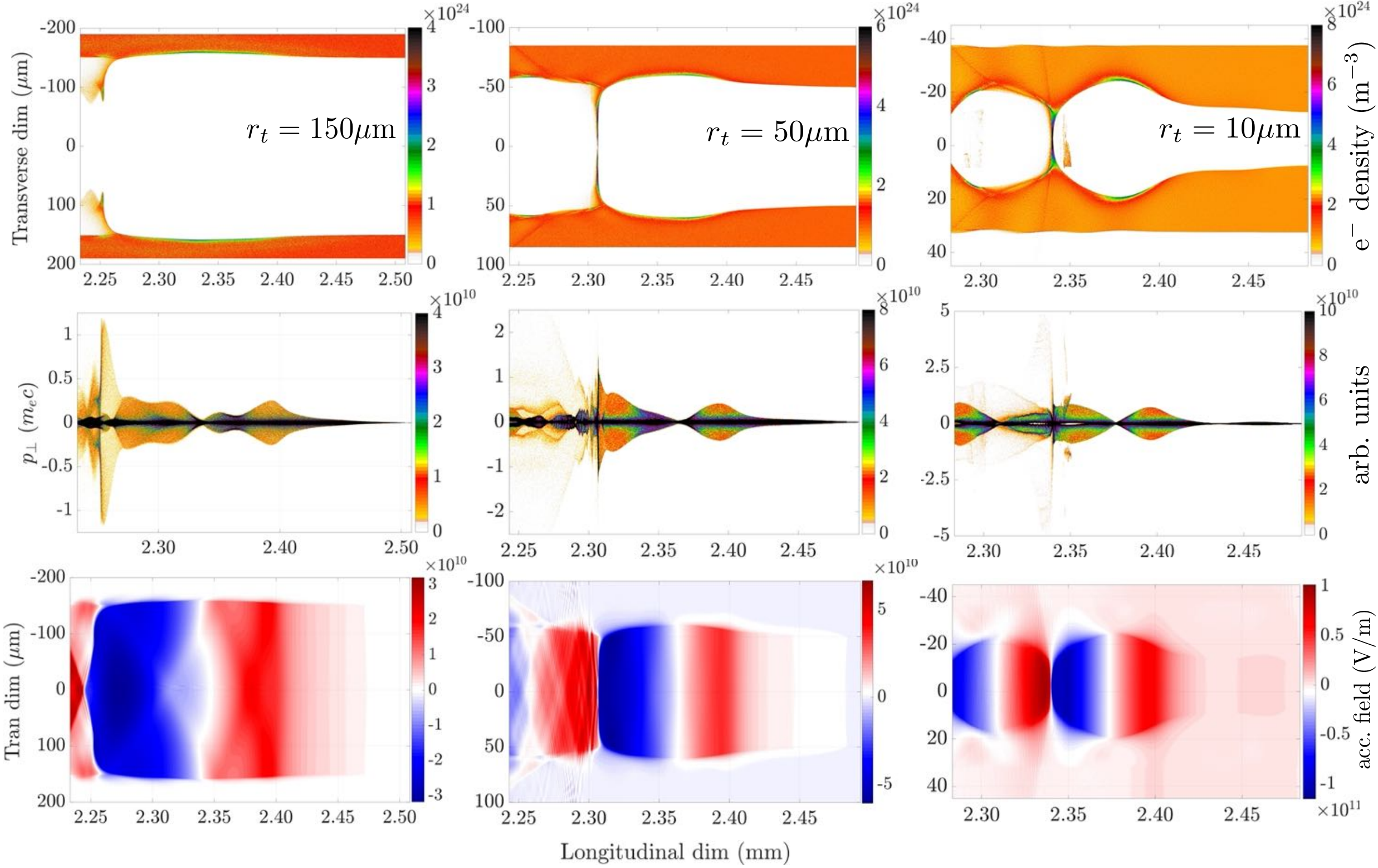}
   \caption{Semiconductor surface crunch-in plasmonic mode excited in tubes of different radius $r_t \rm =150, 50, 10\mu m$ after 2.4mm of interaction. Top row shows the free electron density profile in $\rm m^{-3}$, middle row the transverse momentum vs. longitudinal spatial dimension phase-space (in arbitrary units) and bottom row the longitudinal electric field in V/m.}
\label{fig:facet-ii-run1-tube-radius-scan}
\end{figure*}

\begin{figure*}[!htb]
\centering
   \includegraphics[width=\textwidth]{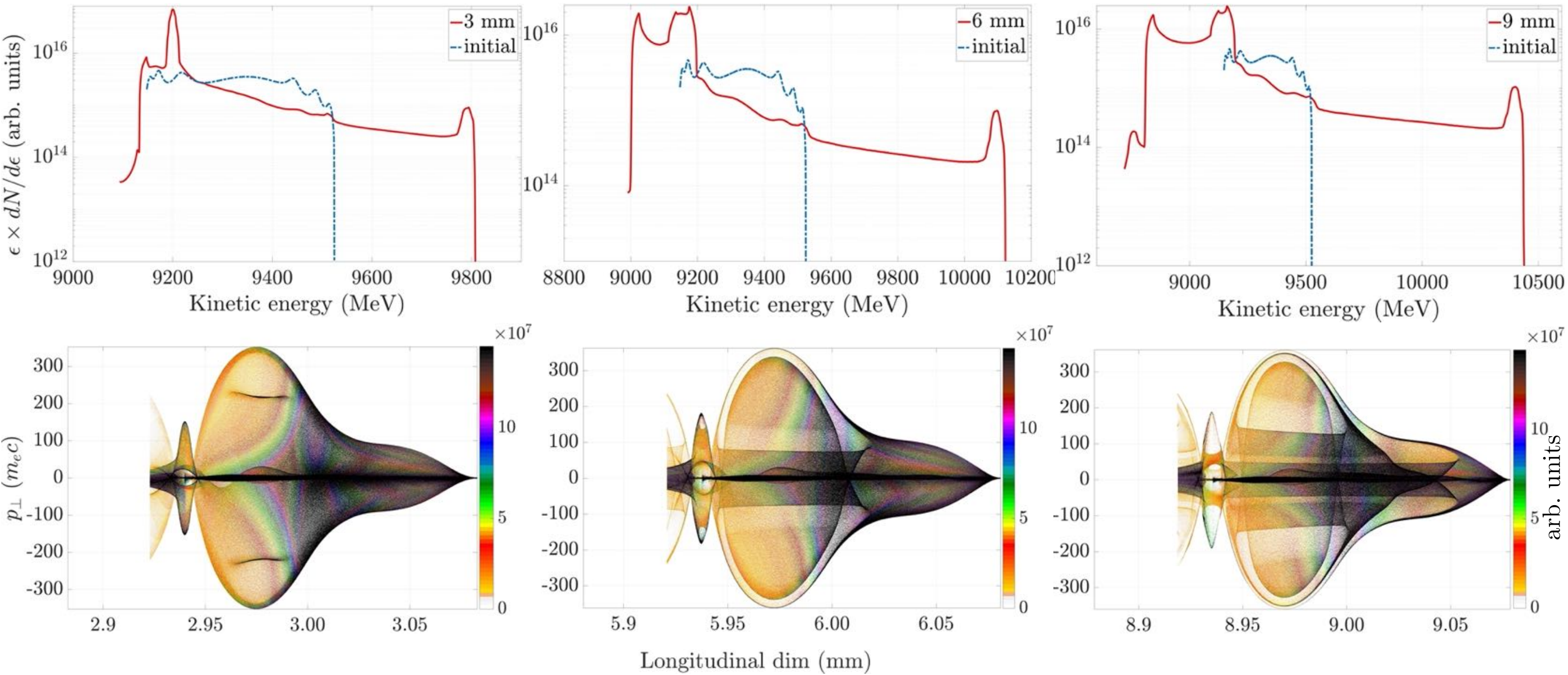}
   \caption{Length scan over 3,6 and 9 mm (left to right column respectively) of beam interaction with plasmonic fields which results in experimentally measurable changes in the beam energy spectra (top row) and transverse phase-space (transverse momentum vs. longitudinal spatial dimension) of the electron beam (bottom row) for $r_t=20\mu\rm m$.}
\vspace{-5.0mm}
\label{fig:facet-ii-run1-effect-on-beam}
\end{figure*}

\vspace{-5.0mm}
\subsection{Scaling of surface crunch-in plasmon with\\ tube dimensions}
\label{tube-radius-scan}

A scan over the radius of semiconductor tube at fixed free electron Fermi gas wall density $n_t = 10^{18}\rm cm^{-3}$ using 2.5D PIC simulations (cartesian grid) is summarized in \noindent Fig.\ref{fig:facet-ii-run1-tube-radius-scan}. It demonstrates matched excitation of semiconductor surface crunch-in plasmon is robust and its properties such as wavelength, Eq.\ref{eq:crunch-in-wavelength} can be further tuned by choosing appropriate tube radius. Corresponding variation of oscillation momentum of the Fermi gas (middle row of Fig.\ref{fig:facet-ii-run1-tube-radius-scan}) and longitudinal field strengths (bottom row of Fig.\ref{fig:facet-ii-run1-tube-radius-scan}) are compared for $r_t = 150, 50, 10\rm\mu m$. 

Firstly, over this range of tube radii the crunch-in mode is consistently excited as the electron gas gains relativistic momentum ($\langle\gamma_e\rangle\gg 1$) and almost instantaneously tunnels through the surface. The mutual electrostatic repulsive force of free electrons that collectively tunnel and accumulate inside the tube provides restoring force which sets up the oscillations. Unlike \cite{plasmonics-ieee-2021}, here the tube encloses nearly the entire beam in the transverse direction with $r_t > \sigma_r$, except for a small fraction in the Gaussian wings that directly interacts with the lattice.

Secondly, radial scan helps verify the kinetic model of plasmon in Eq.\ref{eq:surface-wave-equation} and the parameters derived from it. With decreasing tube radius $r_t$, an increase in maximum radial displacement of electron gas $\Delta r = r_m-r_t$ per Eq.\ref{eq:crunch-in-max-r}, crunch-in wavelength per Eq.\ref{eq:crunch-in-wavelength} and the field strength per Eq.\ref{eq:crunch-in-charge-acc-field-SI} is observed (in proportion with increasing radial momentum gain of electron gas). Similar to 3D PIC simulations, a good agreement is observed between simulations and analytical expressions for various derived quantities. The variation in longitudinal and focusing fields observed for varying tube radius is found to be significant enough to be discernible in experiment as described in sec.\ref{plasmonic-effect-beam}. The first stage of material fabrication makes it possible to study these effects in square semiconductor tubes with 30 and 100$\rm \mu m$ sides.

Importantly, although the beam as modeled in sec.\ref{facet-run-1-beam} is initially misaligned relative to the tube axis with the beam centroid offset in the negative x-direction as seen in Fig.\ref{fig:facet-ii-run1-phase-spaces}(a), but no filamentation or hosing instability is observed in the 3D and 2.5D simulations. This suppression of instabilities is observed over centimeter scale interaction lengths nearly equal to the beam beta function of about 5cm.

Relativistic momentum gain of the free electron Fermi gas is observed over the range of radii modeled in the simulations. Therefore, relativistically induced ballistic transport dominates and Ohm's law cannot be applied. However, the magnitude of relativistic momentum decreases with increasing tube radius and further investigations are required to determine the threshold of transition from ballistic to Ohmic transport.
\vspace{-5.0mm}

\subsection{Observable effects of plasmonic fields}
\label{plasmonic-effect-beam}

Experimentally observable effects of plasmonic fields on the beam are modeled using 2.5D PIC (cartesian grid) simulations. The tube length is varied over a range of the ongoing structured semiconductor fabrication effort (described in sec.\ref{semi-fab}) with a fixed tube radius $r_t = \rm 20\mu m$. 

Focusing and acceleration of electron beam under the action of the fields of crunch-in plasmon is summarized in Fig.\ref{fig:facet-ii-run1-effect-on-beam} using a length scan over 3, 6 and 9 mm long semiconductor tube (left to right columns, respectively).

Fig.\ref{fig:facet-ii-run1-effect-on-beam} (top row) demonstrates the acceleration of a significant fraction of the beam particles by many hundreds of MeV over sub-centimeter tube lengths. Specifically, nearly a GeV energy gain with a quasi-monoenergetic peak at around 10.4GeV (compared to the initial mean beam energy of 9.3GeV observed from Fig.\ref{facet-run-1-beam}(e)) is observed over 9 mm of interaction length. The initial energy spread of the beam is evident in Fig.\ref{facet-run-1-beam}(e). This is in accordance with an average accelerating field of about 100GV/m evident in Fig.\ref{fig:facet-ii-run1-tube-radius-scan} for $r_t=10\mu m$. Note that over longer interaction lengths the particles in the beam wings get focussed into the tube which increases the beam fields at $r \geq r_t$ and as a result the plasmonic fields.

The action of the focusing fields of the surface crunch-in plasmon is experimentally quantifiable based upon the modification of the transverse momentum phase-space of the beam along the longitudinal dimension in the bottom row of Fig.\ref{fig:facet-ii-run1-effect-on-beam}. The changes in this phase-space imprint themselves on spatial modulation of the beam envelope observed downstream from the interaction region. The opening angle observed is around 16 milli-radians ($p_{\perp}/p_{\parallel} = 320/19800 \simeq\rm 16\times 10^{-3} rad$). Moreover, the transverse phase-space has a specific structure due to the structure of the plasmonic focusing fields.

The aforementioned effects of the acceleration and focusing of an electron beam under plasmonic fields can be resolved using existing beam diagnostics.

\vspace{-5.0mm}
\subsection{Semiconductor vs. metallic, dielectric tube}
\label{semi-vs-metal}

\begin{figure}[!htb]
\centering
   \includegraphics[width=0.68\columnwidth]{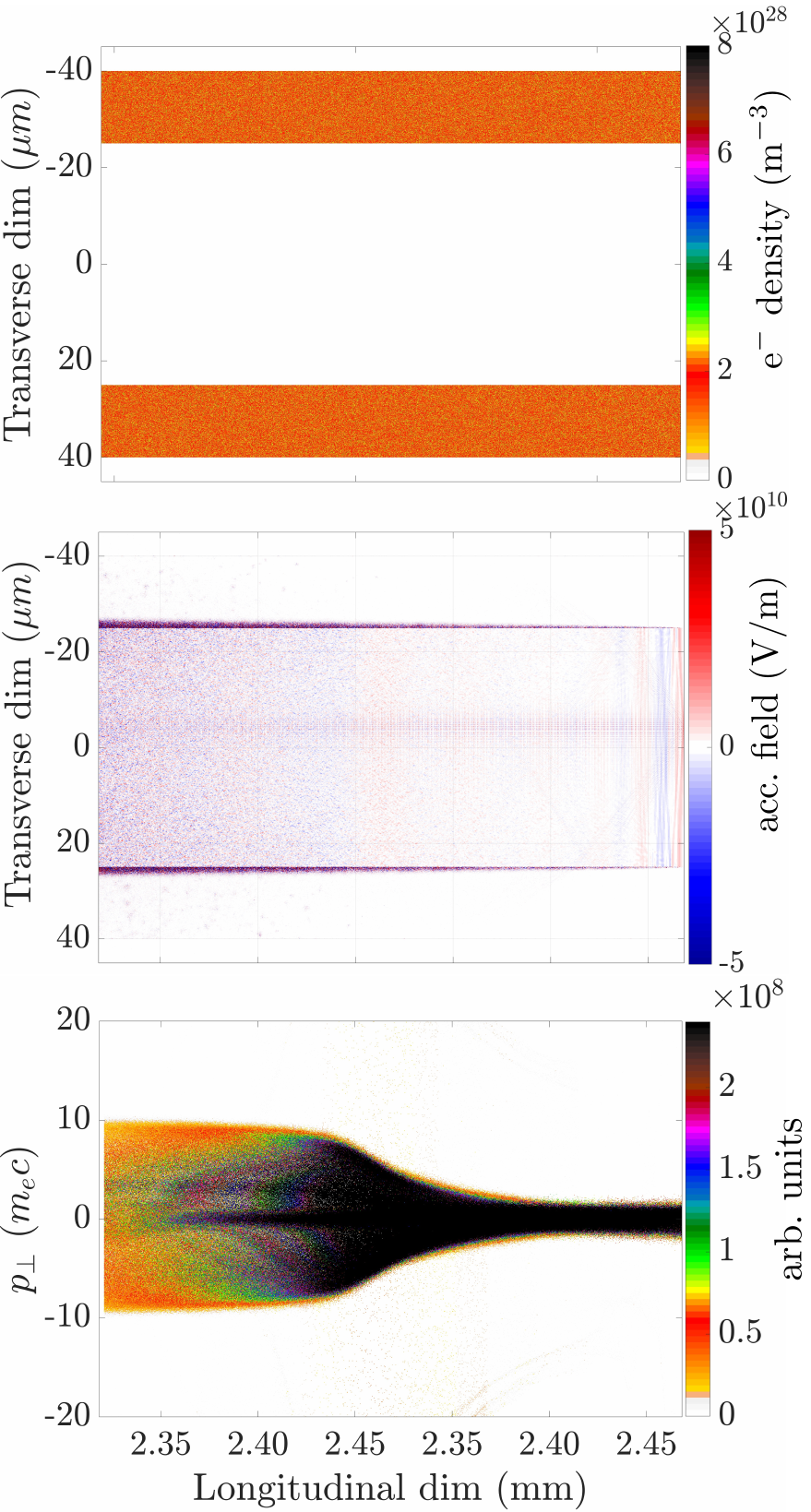}
   \caption{Electron density (top), accelerating field (middle) and beam transverse momentum phase-space (bottom) after 8ps of interaction between a metallic tube with free electron density, $n_t=2\times 10^{22}\rm cm^{-3}$ excited by the beam in sec.\ref{facet-run-1-beam}.}
\vspace{-5.0mm}
\label{fig:facet-ii-run1-semi-vs-metal}
\end{figure}

A critical aspect of plasmonic modes is the ability to tune the conduction band electron density and thereby control the free electron Fermi gas density which makes it possible to match with readily available beams. Here a comparison of the excitation of tubes with walls of semiconductor versus metallic plasmonic structures (modeled in \cite{plasmonics-ieee-2021}) using currently available nC, ten micron electron beam with properties described in sec.\ref{facet-run-1-beam} is used to demonstrate this. While semiconductor material with free electron density of $\rm 10^{18}cm^{-3}$ has a characteristic  plasmon size of $\rm 33\mu m$, for metallic free electron density of $\rm 2\times 10^{22}cm^{-3}$ it is only $\rm 250nm$. The currently accessible beam dimensions of $\rm \sigma_r,\sigma_z\simeq 10\mu m$ and the peak beam density $\rm n_{b0}\simeq 10^{18}cm^{-3}$ are therefore largely mismatched with that of metallic plasmons modeled in \cite{plasmonics-ieee-2021}.

Simulation of plasmonic mode excited by the nC, ten micron electron beam is summarized in Fig.\ref{fig:facet-ii-run1-semi-vs-metal}. The excited free electron density profile (top), acceleration field (middle) and beam transverse phase-space (bottom) is presented after 8ps of interaction.

It is evident that the profile of the acceleration field in metallic density excited under the heavily mismatched condition hardly has any spatial structure or features. Consequently, no net acceleration or deceleration of beam particles is observed unlike for a semiconductor plasmon in the middle row of Fig.\ref{fig:facet-ii-run1-tube-radius-scan}. Similarly, on average the focusing fields have an almost negligible effect on the beam. From the featureless beam transverse phase, the maximum opening angle of beam particles is only $5\times10^{-4}$ rad while the 15.2 milli-rad for the semiconductor plasmonic mode which is experimentally measurable has the imprint of plasmonic focusing fields.

On the other hand, a dielectric tube of similar dimension made of silica ($\rm SiO_2$, an insulator) but without a metallic coating is used for direct comparison with plasmons. Per the discussion in sec.\ref{plasmonic-disambiguation}.a, an uncoated dielectric tube does not trap the Cherenkov radiation and therefore no observable effects on the beam are expected.

\vspace{-5.0mm}
\section{Structured Semiconductor \\ Fabrication}
\label{semi-fab}

The first stage of semiconductor plasmonic structures have been fabricated on Silicon wafers of 100mm diameter polished on a single side. These wafers were $\rm 525 \mu m$ thick with $\langle100\rangle$ orientation and chosen with either high resistivity ($\rm 20 \Omega$-cm) or low resistivity ($\rm 0.025 \Omega$-cm). The n-type doped low resistivity wafer corresponds to a conduction band free electron density of about $\rm 10^{18}cm^{-3}$. The wafers were cleaned using standard procedures, coated with AZ P4330 photoresist and groups of channels 30 and 100 $\rm\mu m$ wide, approximately 1.5mm apart and a range of lengths 3-30mm, were defined using a direct write laser exposure (maskless) lithography. Deep reactive ion etching (DRIE) was used to etch channels of square cross section. A second wafer was processed similarly except that the etching was done on the rough, unpolished side of the wafer.  The polished sides of the two wafers were put in contact in an evacuated chamber and then transferred to an oven where they were annealed at $\rm 1100 ^{\circ} C$ in Nitrogen atmosphere for three hours to form a permanent bond. The etched pattern in the top unpolished surface served as a guide to saw the wafer into pieces with groups of channels of the same length.

\begin{figure}[!htb]
\centering
   \includegraphics[width=0.8\columnwidth]{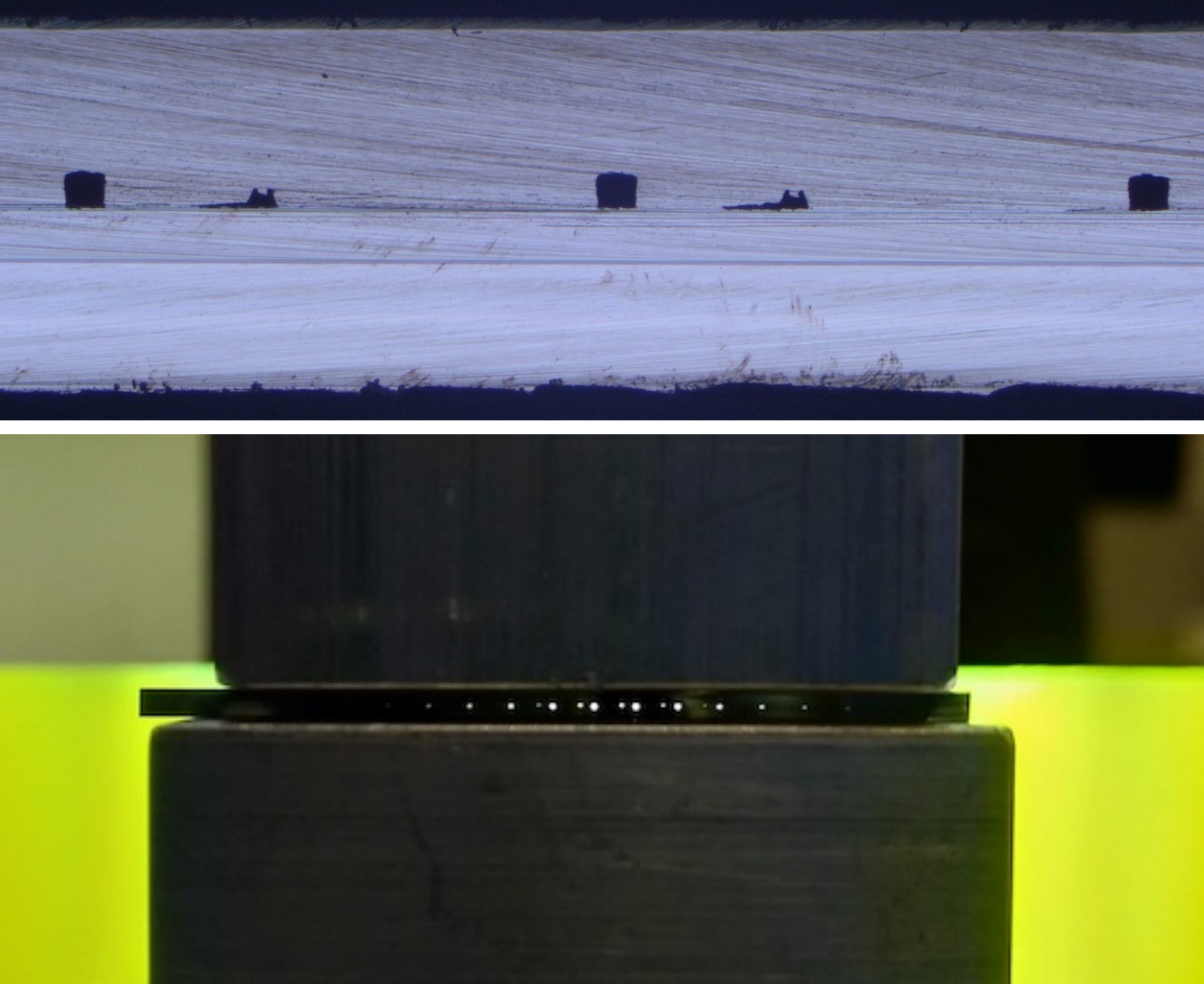}
   \caption{Cross-section of semiconductor tubes (top) with $100\times 100\mu\rm m^2$ and $30\times 30\mu\rm m^2$ cross-sections stacked adjacent to each other in a 3mm end-to-end zoomed-in field-of-view. Light passing through the length of these tubes (bottom).}
\vspace{-2.5mm}
\label{fig:semiconductor-tube-fabrication}
\end{figure} 

The top panel of Fig.\ref{fig:semiconductor-tube-fabrication} is 3mm wide expanded view of the prepared semiconductor sample showing three $\rm 100 \mu m$ and two $\rm 30 \mu m$ channels. The bottom panel is a full end-to-end view of the structure showing light passing through the adjacent pairs of $\rm 100 \mu m$ and $\rm 30 \mu m$ square channels of 3mm length. Relative ease of first aligning the beam with wider channels makes it possible to experimentally converge towards smaller channels that are at a fixed distance from the already aligned wider ones.
\vspace{-5.0mm}

\section{Towards Mega-Ampere currents: \\ sub-micron beams}
\label{MA-beam}

While structured semiconductors are used to match with currently accessible ten micron bunches, the ultimate reach of PV/m plasmonics is attainable in materials with highest free electron densities. However, matched excitation of such metallic plasmons requires sub-micron bunch lengths and waist-sizes which closely align with ongoing trends in bunch compression. This section summarizes ongoing efforts and challenges behind MegaAmpere (MA) peak currents using submicron, nC bunches.

\vspace{-5.0mm}
\subsection{Sub-micron bunch length compression\\ of nC charge beams}
\label{length-comp}
Recent advances in beam phase-space rotation based bunch length compression techniques \cite{acc-handbook} has opened up the possibility of sub-micron bunch lengths with multi-MA peak currents. Matched excitation of plasmons in metals relies on compression of bunch length and focusing of bunch waist to sub-micron dimensions.

Ongoing upgrades at the FACET-II facility of the Stanford Linear Accelerator Center (SLAC) are expected to provide greater than 200kA pulses with less than $\rm 1 \mu m$ rms bunch length in the near future. The precedent to sub-micron bunches was set by $\rm20 ~ to ~ 30 \mu m$ rms bunch length, with 30kA peak current that were in operation at the same facility during the last decade.

Extension of longitudinal compression of electron bunches into the $\rm 100 nm$, multi-MA regime has been modeled in recent feasibility studies \cite{gwhite-design}. 

The regime of multi-MA peak current compression requires next-generation high-brightness electron injector apart from mitigation of various non-linearities present in the compression process to preserve both longitudinal and transverse properties of the bunch such as emittance. 

Extreme compression is limited primarily by coherent synchrotron radiation (CSR) \cite{csr-talman} in various bends which causes transverse emittance growth of many orders of magnitude, in addition to greatly limiting the final achievable peak current. To compensate for emittance degradation due to CSR, more complex compression chicane designs beyond the standard 4-bend chicane have been considered, e.g. multi-bend chicanes, quadrupole and sextupole loaded chicanes, arcs or wigglers.

\noindent The key design parameters, constraints and considerations for a bunch compressor capable of compression of electron bunch to peak currents $> \rm 0.3 MA$ (1nC bunch compressed to $\rm 1 \mu m$ bunch length) with $< \rm 1mm-mrad$ transverse emittance growth are:
\begin{enumerate}[topsep=0pt, itemsep=0.3ex, partopsep=0.3ex, parsep=0.3ex,leftmargin=*]
\item an advanced photo-injector with transverse emittance $<1 \rm mm-mrad$, 1-2 nC charge, few 100A initial peak current ($\sim\rm 3mm$ initial bunch length at the injector) and typical injector energy of $\sim 100 \rm MeV$.
\item initial injector charge $\rm > 2nC$ to allow for collimation of tails or wings formed during compression.
\item	multi-stage compression to achieve bunch length compression ratios of $\rm >1000$ as stage-wise compression removes the need for excessive energy spread to achieve such large compression ratios. For example, FACET-II uses three compression stages at 0.3, 4.5, 10 GeV beam energy along the linear accelerator.
\item final compression needs to be at the final beam energy to account for the following effects:
	\begin{itemize}[topsep=0pt, itemsep=0.3ex, partopsep=0.3ex, parsep=0.3ex,leftmargin=*]
		\item relative emittance growth due to CSR decreases with beam energy as $\Delta \epsilon_n/\epsilon_n \sim 1/\sqrt{\gamma}$
		\item adverse surface effects in accelerator components are expected at the highest compression (ohmic pulse heating) and the final compression stage minimizes the interaction at maximum compression
		\item energy chirp induced by cavity wakefields in the final accelerating section is utilized. The final compression stage should have positive longitudinal momentum to space correlation parameter $\rm R_{56} > 0$, with higher energy particles at the head of the bunch, to utilize cavity wakefield chirp. This also helps with non-linearities from compression system ($\rm T_{566}$) with partial self-cancelation from rf curvature.
	\end{itemize}
\end{enumerate}


Initial compression stages reduce the bunch length to $\rm 20\mu m$ rms, where the peak current is a 12 kA, similar to the largest values used at FEL facilities, where emittance preservation has been experimentally demonstrated at low charge ($\rm < 200 pC$).

To achieve the desired positive $\rm R_{56}$ compression for the final compression stage a quadrupole-loaded lattice such as an arc or wiggler (depending on whether net-bending is a desired feature or not) is required. 

A proof-of-principle model using the tracking code Lucretia \cite{lucretia-code} utilized a wiggler system designed to perform the tasks of the final compressor. This design uses a triplet-based arc wiggler (9 cells) with a total bend angle of 67 mrad at 30 GeV. This provides +17mm of $\rm R_{56}$ in a length of 150m whilst keeping emittance growth to $< 5\%$ due to incoherent synchrotron radiation in the bends. Sextupoles are included in the optics to provide cancelation of CSR energy kicks. The maximum compression for this design is limited at 375 kA, with 5 mm-mrad of slice emittance growth. Further work is ongoing to control emittance growth and achieve higher peak currents.

\vspace{-5.0mm}
\subsection{Sub-micron waist-size compression with \\ Permanent Magnet Quadrupoles (PMQs)}
\label{waist-comp}

Excitation of the strongly electrostatic surface crunch-in plasmonic mode requires beam densities on the order of $\rm 10^{17}cm^{-3}$ to $\rm 10^{21} cm^{-3}$. Besides high density, the beam dimensions need to approach plasmonic wavelength in Eq.\ref{eq:plasmonic-wavelength}. Both these correspond to sub-micron transverse spot sizes to excite metallic densities. In order to produce conditions optimal for matched excitation of the crunch-in mode in metallic density materials, an appropriate focusing scheme is required. Furthermore, in realistic experimental scenarios where interaction regions are constrained in space, the focusing system should be compact and robust. A magnetic focusing system that can work along with plasmonic nanofocusing \cite{plasmonic-nanofocusing} is modeled. 

\begin{figure}[h]
\centering
   \includegraphics[width=\columnwidth]{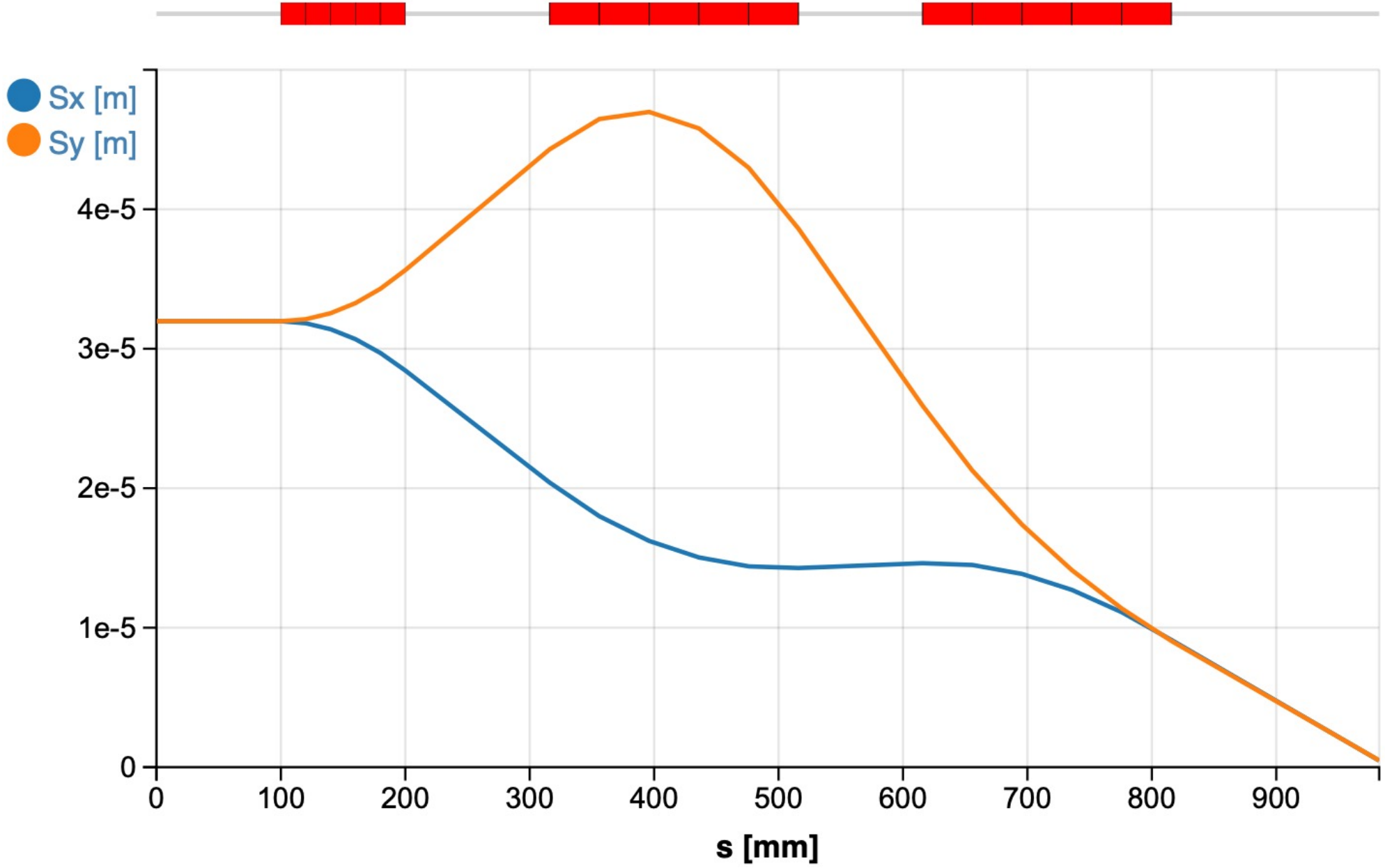}
   \caption{Particle-tracking simulation \cite{elegant-code} demonstrating the compression of an electron beam initially with ten micron waist-size $\rm\sigma_{x,y}\sim 10\mu m$ to $\rm<500nm$ in a $\rm 1 m$-long PMQ by first defocusing it to about $\rm 30\mu m$ coming into the PMQ.}
\vspace{-5.0mm}
\label{fig:PMQ-triplet}
\end{figure}

PMQs are ideal focusing elements for small spot size applications as they are able to provide very high magnetic field gradients. These gradients can be in excess of 700 T/m \cite{adj-pmq} for rare-earth magnets and up to 3000T/m in  nanofabricated designs \cite{nanomaterial-pmq} in a compact form factor. However, quadrupoles only provide focusing in one dimension, while defocusing the beam in the other. Hence, the simplest configuration of permanent magnet quadrupoles for overall focusing requires a triplet configuration. The triplet system is tunable by the relative reconfiguration of quadrupole locations. Hybrid configurations, with steel, can be used to incorporate split designs for limited access. 

For the present configuration, consideration of the limited space upstream of the structure is critical such that both high intensity, and adequate interaction length is achievable. So, only configurations constrained to $\rm <1m$ are considered.  In order to attain this goal, permanent magnet quadrupoles with gradient $\rm\sim 750T/m$ are used in a triplet layout shown as red blocks in Fig.\ref{fig:PMQ-triplet}, with lengths of 10cm, 20cm, and 20cm respectively. The final spot size using this triplet is just below 500nm in both the transverse dimensions. A defocusing arrangement downstream of the focal locations with mirror symmetry would transport the spent beam after the interaction.

%
%

\section{Conclusion}
\label{conclusion}

The tunability of Fermi electron gas in structured semiconductors allows matched excitation of large-amplitude surface plasmons using a range of currently accessible electron beams. It is computationally demonstrated that by using n-typed doped semiconductors with free electron Fermi gas density of about $\rm 10^{18}cm^{-3}$ the plasmonic mode dimensions match currently available ten micron, nC electron beam. This makes it possible to undertake immediate  experimental verification of the underlying principles of Petavolts per meter plasmonics, including relativistic ballistic transport and relativistic tunneling of electron oscillations that traverse the surface.

Excitation of large-amplitude, surface crunch-in plasmonic mode in semiconductor tubes is modeled to sustain tens of GV/m acceleration and focusing fields inside the tube. Such large plasmonic fields not only accelerate a significant number of beam particles to gain hundreds of MeV in sub-centimeter long tubes but also focus the beam resulting in tens of milli-radian angular deviation. These effects can be resolved using available diagnostics. Metallic and dielectric tubes of similar dimensions do not produce these effects.

Experimental verification of the underlying details of Petavolts per meter plasmonics using tunable semiconductor structures hand in hand with the ongoing MegaAmpere sub-micron beam effort can open transformative pathways in fundamental science and technological applications at the froniters of electromagnetism.

\vspace{2.5mm}
\begin{acknowledgments}
This work was supported by the Department of Electrical Engineering at the University of Colorado Denver. The NSF XSEDE RMACC Summit supercomputer utilized for computational modeling was supported by the NSF awards ACI-1548562, ACI-1532235 and ACI-1532236, the Univ. of Colorado Boulder, and Colorado State Univ. \cite{xsede-rmacc-citation-1,xsede-rmacc-citation-2}.
\end{acknowledgments}


\end{document}